\documentclass[a4paper,fleqn,usenatbib]{mnras}

\usepackage{newtxtext,newtxmath}

\usepackage[T1]{fontenc}
\usepackage{tabularx}

\DeclareRobustCommand{\VAN}[3]{#2}
\let\VANthebibliography\thebibliography
\def\thebibliography{\DeclareRobustCommand{\VAN}[3]{##3}\VANthebibliography}

\usepackage{graphicx}	
\usepackage{amsmath}	
\usepackage{subfig}
\usepackage{multirow}
\usepackage{multicol}
\DeclareUnicodeCharacter{2212}{-}

\title[]{The PAU Survey: Uncovering the connection between intrinsic and observed galaxy properties using symbolic regression}

\author[Kumar et al.]{Adarsh Kumar,$^{1,2}$\thanks{E-mail: adarsh.kumar.25@alumnos.uda.cl}
Carlton M. Baugh,$^{3,4}$
Suttikoon Koonkor,$^{3,4,5}$ 
Giorgio Manzoni,$^{6}$ 
Sukanta Panda,$^{1}$\newauthor
D.~Navarro-Giron\'es,$^{9}$
R.~Casas,$^{7,8}$
J.~Carretero,$^{12, 13}$
F.~Castander$^{8}$,
J.~De Vicente,$^{12}$
J.~Garc\'ia-Bellido,$^{11}$\newauthor
E.~Gaztañaga,$^{10,7,8}$
R.~Miquel,$^{14,15}$
P.~Renard,$^{7,8}$
P.~Tallada-Cresp\'{i}$^{12,13}$
\\
\\
\\
$^{1}$Department of Physics, Indian Institute of Science Education and Research, Bhopal, 462066, India.\\
$^{2}$Instituto de Astronomía y Ciencias Planetarias, Universidad de Atacama, Copayapu 485, Copiapó, Chile\\
$^{3}$Institute for Computational Cosmology, Department of Physics, Durham University, South Road, Durham DH1 3LE, UK.\\ 
$^{4}$Institute for Data Science, Durham University, South Road, Durham DH1 3LE, UK\\
$^{5}$National Astronomical Research Institute of Thailand, Don Kaeo, Mae Rim, Chiang Mai, 50180, Thailand\\
$^{6}$State Key Laboratory of Displays and Opto-Electronics, The Hong Kong University of Science and Technology, Hong Kong S.A.R., China.\\
$^{7}$Institute of Space Sciences (ICE, CSIC), Campus UAB, Carrer de Can Magrans, s/n, 08193 Barcelona, Spain\\
$^{8}$Institut d'Estudis Espacials de Catalunya (IEEC), E-08860 Castelldefels (Barcelona), Spain\\
$^{9}$Leiden Observatory, Leiden University, Einsteinweg 55, 2333 CC Leiden, the Netherlands\\
$^{10}$Institute of Cosmology \& Gravitation, University of Portsmouth, Dennis Sciama Building, Burnaby Road, Portsmouth PO1 3FX, UK\\
$^{11}$Instituto de F\'isica Te\'orica CSIC/UAM, Universidad Aut\'onoma de Madrid, 28049 Madrid, Spain \\
$^{12}$Centro de Investigaciones Energéticas, Medioambientales y Tecnológicas (CIEMAT), Avenida Complutense 40, 28040 Madrid, Spain.\\
$^{13}$Port d’Informaci\'o Cient\'ifica (PIC), Campus UAB, C. Albareda s/n, 08193 Bellaterra (Barcelona), Spain.\\
$^{14}$Institut de F\'isica d’Altes Energies (IFAE), The Barcelona Institute of Science and Technology, Campus UAB, 08193 Bellaterra (Barcelona), Spain.\\
$^{15}$Instituci\'o Catalana de Recerca i Estudis Avan\c{c}ats (ICREA), 08010 Barcelona, Spain.
\\}

\date{Accepted XXX. Received YYY; in original form ZZZ}

\pubyear{2025}

\begin{document}
\label{firstpage}
\pagerange{\pageref{firstpage}--\pageref{lastpage}}
\maketitle


\begin{abstract}
Estimating stellar masses for billions of galaxies in upcoming surveys requires methods that are both accurate and computationally efficient. We present a new approach using symbolic regression trained on a simulation to derive simple, explicit mathematical expressions that estimate galaxy stellar masses from basic observables: photometry and redshift. Using a mock catalogue from the \texttt{GALFORM} semi-analytical model that reproduces the Physics of the Accelerating Universe Survey (PAUS), we show that a linear combination of just four observables—minimally processed $u$- and $i$-band magnitudes, observed $(g-r)$ colour, and redshift—can recover stellar masses with accuracy comparable to traditional spectral energy distribution (SED) fitting, but with negligible computational cost. Our expressions can be evaluated instantaneously for millions of galaxies, making them ideal for next-generation surveys like LSST and Euclid. When observational errors are included, symbolic regression achieves a similar accuracy to deep neural networks while maintaining transparency. Validation against \texttt{CIGALE} SED fitting on PAUS data shows agreement within 0.13~dex for galaxies with $M_* > 10^8 M_\odot$. We demonstrate that the stellar mass function can be recovered at $z < 0.5$, though with distortions at the extremes: the high-mass end is overestimated by a factor of $\sim$3 at $10^{11.5} h^{-1}M_\odot$ due to scatter. Our approach offers a fast, transparent alternative to traditional methods without sacrificing accuracy for the bulk of the galaxy population.
\end{abstract}




\section{Introduction}

With the increasing amount of galaxy photometry becoming available from wide-field surveys, it is crucial to find fast and accurate ways to determine the intrinsic properties of galaxies, such as stellar mass, from observables. The Dark Energy Spectroscopic Instrument (DESI) Legacy Imaging Surveys cover 14,000 square degrees in several optical bands \citep{Dey2019}, while the upcoming Rubin Observatory Legacy Survey of Space and Time will image over 18,000 square degrees in six optical filters \citep{Ivezic2019ApJ}. These surveys will detect billions of galaxies, making traditional analysis methods computationally challenging. Stellar mass is a fundamental property that connects observations to the underlying physics of galaxy formation and evolution, serving as a key diagnostic for testing theoretical models and understanding the galaxy population across cosmic time.

The problem of inferring intrinsic galaxy properties from observables is typically solved using forward modelling through spectral energy distribution (SED) fitting. In this approach, a star formation history is combined with a stellar population synthesis code to produce a composite spectral energy distribution for a galaxy, which is then matched to observed photometry through template fitting \citep{Bouquien2019, Robotham2020, Johnson2021}. This methodology has been widely and successfully applied across a range of surveys and redshifts. \citet{Salim2016} applied this approach to estimate the star formation rates, stellar masses, and dust attenuations for 700,000 galaxies with redshift $z < 0.3$ from the Sloan Digital Sky Survey. \citet{Siudek2024} used a similar methodology to create a value-added catalogue for 1.3 million galaxies from the DESI surveys. \citet{Wang2024} applied the method to a sample covering a very wide redshift range using Hubble Space Telescope (HST)\footnote{\url{https://science.nasa.gov/mission/hubble/}} and James Webb Space Telescope (JWST)\footnote{\url{https://science.nasa.gov/mission/webb/}} photometry. Using the Physics of the Accelerating Universe Survey (PAUS), \citet{Csizi2024} investigated whether supplementing broad-band photometry with narrow-band imaging in many bands improved the accuracy of stellar mass estimates.

However, the spectral synthesis approach involves significant uncertainties and modelling choices. Key assumptions include the form of the star formation history, the treatment of stellar metallicity evolution, and the amount of attenuation of starlight by dust \citep{Pforr2012, Mitchell2013}. \citet{Zhang2017} showed that holding the metallicity fixed or varying the dust attenuation can lead to biases of a factor of two or more in the recovered stellar masses. \citet{Bowman2020} analysed HST CANDELS galaxies and found that the form of the star-forming sequence at low masses depends strongly on the assumed star formation history, with estimated stellar masses varying by an order of magnitude in different star formation scenarios. These systematic uncertainties are difficult to quantify and can dominate over statistical errors, particularly for galaxies with complex star formation histories or significant dust obscuration. Furthermore, SED fitting is computationally expensive, requiring iterative optimization over large template libraries for each galaxy—a process that becomes prohibitive for the billions of galaxies expected from next-generation surveys.

To address the computational challenge, several authors have explored using machine learning methods to accelerate stellar mass estimation. \citet{Guo2024} used an artificial neural network to predict stellar masses and star formation rates using broad-band photometry as input, training on spectral energy distributions generated using \texttt{CIGALE}. \citet{Zeraatgari2024} employed decision trees and deep learning architectures to learn the connection between stellar mass and observables from a value-added catalogue, demonstrating that stellar masses can be recovered with typical scatter of 0.2~dex for low-redshift galaxies. Several studies have investigated applying machine learning to Euclid\footnote{\url{https://www.esa.int/Science_Exploration/Space_Science/Euclid}} data to infer galaxy properties including stellar mass \citep{Bisigello:2023} and to classify galaxies \citep{Humphrey2023}. These machine learning approaches typically learn relationships from catalogues where stellar masses have been derived through traditional SED fitting, effectively interpolating within the assumptions and systematics of the original method while achieving substantial computational speedup.

An alternative approach uses physical galaxy formation models to generate ensembles of synthetic galaxies with known properties, which can then be used to train machine learning models without the intermediate step of SED fitting. \citet{Simet2021} used a semi-analytical galaxy formation model, adjusted to match the CANDELS survey, to train a neural network to estimate intrinsic galaxy properties directly from photometry. \citet{Davidzon2019} used the output of a hydrodynamical simulation of galaxy formation to demonstrate that a self-organized map could be built using observables, which grouped galaxies with similar stellar masses. This map was then applied to estimate the masses of real galaxies in \citet{Davidzon2022}. \citet{Santos2024} used a hydrodynamical simulation to train a neural network to extract galaxy redshifts and stellar masses (amongst other properties) using photometry taken with JWST. These approaches have the advantage that they avoid the assumptions inherent in SED fitting and instead learn relationships directly from the underlying galaxy formation physics, though they inherit different systematic uncertainties from the galaxy formation models themselves.

Here, we extend this simulation-based approach by using symbolic regression to derive explicit mathematical expressions that relate stellar mass to observable quantities. Unlike neural networks or other ``black box'' machine learning methods, symbolic regression seeks to find transparent functional forms that can be written down as simple equations. This provides both computational efficiency and physical interpretability: the resulting expressions can be evaluated instantaneously without specialized software, and the coefficients reveal which observables drive the stellar mass estimates and how. We use a mock catalogue generated by \citet{Manzoni:2024} from the \texttt{GALFORM} semi-analytical galaxy formation model \citep{Cole:2000, Lacey:2016, Baugh:2019} to train our models. The mock reproduces the Physics of the Accelerating Universe Survey (PAUS), which combines deep narrow-band imaging with broad-band photometry to achieve high-precision photometric redshifts \citep{Eriksen:2019, Serrano:2023}.

We first establish a benchmark using deep learning—specifically, an artificial neural network—to learn the relationship between stellar mass and observables (redshift and broad-band photometry). We apply minimal processing to the observables, avoiding the need to calculate $k$-corrections explicitly. We also investigate how this relationship is affected by including realistic measurement errors in the photometry and photometric redshifts. The deep learning approach should provide the best possible estimation of stellar mass given the observables used and the training sample size, because the multi-layer structure and non-linear activation functions allow complex relationships between inputs and outputs to be captured. However, this richness comes at the cost of transparency: we gain no insight into how the observables affect the stellar mass estimate.

To address this limitation, we apply a range of symbolic regression methods to learn the relationship between stellar mass and observables in the form of explicit equations. These methods search for mathematical expressions of varying complexity that best reproduce the stellar mass in terms of the input observables. We deliberately constrain the complexity of the expressions, motivated by the desire for simple, interpretable formulae rather than ones that allow all possible combinations of terms. This represents a trade-off: we sacrifice some accuracy in exchange for transparency and ease of use. Remarkably, we find that the expressions we derive are similar in form to the physically motivated stellar mass estimators proposed by \citet{Robotham2020}, providing validation that our empirically derived relationships capture the correct underlying physics.

Symbolic regression methods have recently been applied successfully to a variety of problems in astronomy, including galaxy clustering \citep{Miguel:2021, Miguel:2023}, modified gravity \citep{Desmond:2023}, halo mass functions \citep{Bartlett:2024}, and cosmological parameter estimation \citep{Sui:2024}. These applications demonstrate that symbolic regression can discover physically meaningful relationships from complex datasets while maintaining interpretability. Our application to stellar mass estimation represents a natural extension of these techniques to galaxy property derivation.

The layout of the paper is as follows. In Section~2 we describe the galaxy formation model (\S~2.1) and the observational selection of the mock catalogue (\S~2.2). 
In Section~3 we use a deep learning approach to learn the relationship between observables and stellar mass. 
Symbolic regression is applied to the mock catalogue in Section~4, where we compare different regression techniques and derive our best-fitting expressions. We compare the stellar mass predictions from symbolic regression with results from SED fitting in Section~5. The symbolic regression equations are applied to the PAUS data in Section~6 to provide estimates of the stellar mass function. Our conclusions are given in Section~7. The cosmology adopted here matches that in the P-Millennium simulation used to build the mock catalogues \citep{Baugh:2019}: a flat universe with present-day matter density parameter $\Omega_{\rm M} = 0.307$ and reduced Hubble parameter $h = 0.6777$.

\section{Model galaxies}

The data used here was generated using the \texttt{GALFORM} galaxy formation model \citep{Cole:2000}, and is in the form of a mock catalogue of the Physics of the Accelerating Universe Survey (PAUS; \citealt{Serrano:2023},\citealt{Eriksen:2019}) as described in \cite{Manzoni:2024}. This choice gives us access to the photometry predicted for a large number of galaxies, along with the stellar masses of the galaxies, as calculated by the model. Also, we can isolate the effect of observational measurement errors in the photometry of the galaxies and their photometric redshifts on the accuracy of the predicted stellar mass. The \texttt{GALFORM} model is briefly introduced in \S~\ref{sec:galform}. The mock catalogue used here is described in \S~\ref{sec:mock}. The simple models used to add photometric errors and photometric redshift errors are described in \S~\ref{sec:err}. A basic characterisation of the mock PAUS catalogue is presented in \S~\ref{sec:char}; further properties of the mock can be found in \cite{Manzoni:2024}.

\subsection{The galaxy formation model}
\label{sec:galform}

\texttt{GALFORM} is a physically motivated galaxy formation model  \citep{Cole:2000, Bower:2006, Lacey:2016}. \texttt{GALFORM} is used to predict the galactic content of dark matter (DM) haloes at the earliest branches of the halo merger tree with hot baryonic gas and models the key physical processes behind the formation and evolution of galaxies using a set of coupled differential equations. These processes include: (i) the collapse and merging of DM haloes, (ii) the shock-heating and radiative cooling of gas inside DM haloes, leading to the formation of galactic discs, (iii) quiescent star formation in galactic discs, (iv) feedback from supernovae (SNe), active galactic nuclei (AGN), and photo-ionization of the inter-galactic medium, (v) chemical enrichment of stars and gas and (iv) dynamical friction which leads to mergers of galaxies within DM haloes, forming spheroids and triggering starbursts. Full descriptions of these physical processes are given in \citet{Lacey:2016} (see also  \citealt{Baugh:2006} and \citealt{Benson:2010}). 

We use the variant of \texttt{GALFORM} introduced by \cite{Lacey:2016} and re-calibrated for the P-Millennium $N$-body simulation by \cite{Baugh:2019}. The P-Millennium box length is $542.16 h^{-1}$Mpc and the halo mass resolution is $2.12 \times 10^{9} h^{-1} M_{\odot}$. The production of DM halo merger histories for \texttt{GALFORM} is described in \cite{Jiang:2014}. The cosmology and resolution are different in the P-Millennium simulation compared with those used by \cite{Lacey:2016}, hence the need for the small parameter recalibration described in \cite{Baugh:2019}. 

\texttt{GALFORM} predicts the chemical evolution of the gas and stars in galaxies, along with the size of the disc and bulge components and their star formation histories. The model outputs the mass-to-light ratios in a list of filters that are specified at run time. Along with the model for attenuation of stellar emission by dust described in \cite{Cole:2000} (see also \citealt{Lacey:2016}), this allows the model to predict the magnitude of the modelled galaxies in these bands. Critically for this paper, the model predicts the stellar mass of galaxies.

\texttt{GALFORM} is implemented on the halo merger histories extracted from the P-Millennium $N$-body simulation, to give predictions for the galaxy population at the output redshifts or snapshots of the simulation. The P-Millennium simulation has 271 outputs.

\subsection{The mock catalogue}
\label{sec:mock}

The PAUS mock catalogue covers approximately $100 \,\rm{deg}^2$, with a magnitude limit of $i_{\rm AB} = 24$.  \cite{Manzoni:2024} used P-Millennium snapshots in the redshift range $0<z<2$. For some applications, we impose a magnitude limit to the mock of $i_{\rm AB}=22.5$.

PAUS has imaged patches of the sky using 40 narrow-band filters, each with a full-width half maximum of $130$ \AA, with centres spaced at 100 \AA \, over the wavelength range $4500$ \AA \, to  $8500$ \AA. Here we are interested in the PAUS observations that overlap with the W1 and W3 fields of the Canada-France-Hawaii Telescope Lensing Survey (CFHTLenS) which contributes broad-band photometry in the 
 $u^{*}, g, r, i, z$ bands \citep{Erben:2013}. The PAUS narrow-band imaging, when combined with the CFHTLenS broad bands, allows high accuracy photometric redshifts to be estimated, with a centralised scatter of $\sigma_{68}/(1 + z) = 0.02$ to a depth of $i_{\rm AB} = 23$ (\citealt{Navarro:2024}; see also \citealt{Eriksen:2019,Eriksen:2020, Alarcon:2021,Cabayol:2021,Daza:2024}).

\texttt{GALFORM} outputs the properties and positions of the galaxy population in the simulation box at a discrete set of redshift outputs. The \textit{lightcone} is built by interpolating galaxy magnitudes and positions between the values at these discrete redshifts, using the redshift at which the galaxy crosses the observer's lightcone. To build the PAUS mock, \cite{Manzoni:2024} followed the procedure described in \cite{merson13}.
An observer was placed at a position inside the simulation box, and a line-of-sight direction was chosen for the mid-point of the survey. \cite{Manzoni:2024} generated a 100 square-degree patch on the sky, over the redshift range $0<z<2$. This required replication in space of the simulation box to fill the observed volume.

A galaxy crosses the past lightcone of the observer in between two of the simulation output redshifts or snapshots. The positions of the galaxy in the two snapshots are used to estimate its position at the lightcone crossing. \cite{merson13} applied different interpolation procedures for central and satellite galaxies. Central galaxies are assumed to be at the centre of mass of the host DM halo and hence track its motion between the snapshots: here, a linear interpolation is sufficient. Satellite galaxies, however, follow more complicated paths and can enter the observer's past lightcone either before or after their associated central. For this reason, a more sophisticated treatment is used to compute the position of a satellite galaxy, taking into account its orbit around the central (see fig.~2 of \citealt{merson13}). 

Assigning properties to galaxies as they cross the observer's past lightcone using a simple interpolation between snapshots could lead to inaccuracies. 
Manzoni et~al. followed \cite{merson13} and retained the galaxy properties from the snapshot redshift just above the redshift of lightcone crossing (see \citealt{kitzbichler07}). 
The one exception to this is the magnitude of the galaxy in the pre-specified filters in the observer frame. 
The definition of the observer frame depends on redshift and so is slightly different at the two output or snapshot redshifts that straddle the lightcone crossing redshift. A linear interpolation is performed between these two versions of the observer frame magnitudes to compute the observed magnitude at the redshift of lightcone crossing. 
In addition to the band shifting of the observer frame, the luminosity distance that corresponds to the lightcone crossing redshift is used to compute the apparent magnitude of the galaxy in the mock. 
This approach does not take into account any change in the spectral energy distribution of the galaxy between the higher redshift snapshot and the lightcone crossing redshift. 
However, the resulting colour-redshift relation is smooth and contains no trace of the locations of the simulation snapshots, as shown in fig.~A1 in Appendix~A of Manzoni et~al. 

\subsubsection{Simulating observational errors}
\label{sec:err}

We also consider a mock catalogue in which the galaxy magnitudes have been perturbed to include measurement errors and the photometric redshifts mimic the estimation errors expected when applying a photometric redshift code.  
Following the treatment described in \cite{Manzoni:2024}, photometry  
errors are assumed to have a Gaussian distribution in magnitude. The perturbed magnitude in the band labeled by $j$, $m_{j}^{\rm obs}$ is obtained by adding a Gaussian distributed quantity, $x$, which has zero mean and a variance of ${\sigma}_{j}$, to the true magnitude predicted by \texttt{GALFORM}, $m_{j}^{\rm true}$:

\begin{equation}
    m_{j}^{\rm obs} = m_{j}^{\rm true} + x.
\end{equation}

The variance of the Gaussian is related to the signal-to-noise ratio in band $j$, $(S/N)_{j}$ by:

\begin{equation}
    {\sigma}_{j} = \frac{2.5}{\ln 10}\frac{1}{(S/N)_{j}}.
\end{equation}
 The signal-to-noise ratio is set to be 5 at the magnitude limit in a given band. The broad band and narrow band magnitude limits for the PAUS mock are given in \cite{Manzoni:2024}. This model assumes that all galaxies are treated as point sources.

\cite{Manzoni:2024} showed that these photometry errors led to similar photometric redshift errors to those estimated for the PAUS data when fed through a photometric redshift code. To reduce the computation time, this exercise was carried out in Manzoni et
al. for a random sample of the mock galaxies, giving the distribution of estimated photometric redshifts as a function of the true redshift. We sample this distribution using a running bin in  the true redshift to estimate the photometric redshift for all galaxies in the mock. The bin width is chosen so that it is narrow enough so  that the error distribution is well quantified. We have tested that, this way, we can recover the scatter and outlier fraction obtained when running the photometric redshift code directly.

\subsubsection{Characterization of the mock}
\label{sec:char}

\begin{figure}
    \centering
\includegraphics[trim= 1.5cm 2.5cm 3cm 1cm, width=1.1\columnwidth]
{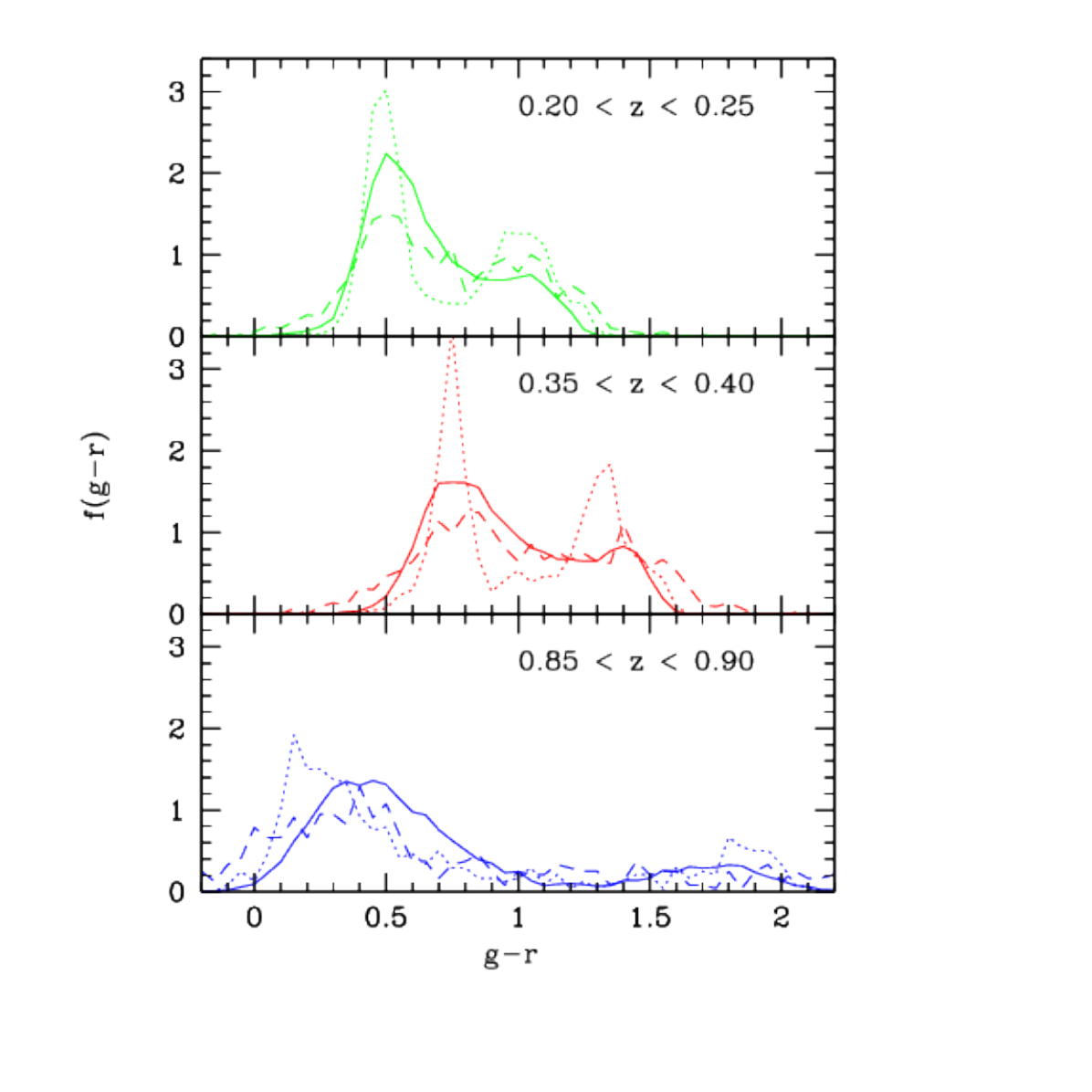}
    \caption{The distribution of observed $g-r$ colour for galaxies with $i_{\rm AB} < 22.5$ in selected redshift slices: top   $0.20<z<0.25$, middle  $0.35<z<0.40$, bottom  $0.85<z<0.90$. The solid lines show the PAUS observations, the dotted lines show the \texttt{GALFORM} mock without photometry errors and the dashed lines show the mock with photometry errors included. The curves have been normalised to enclose the same area. The line colours have no meaning, and have been chosen for artistic merit. 
     }
    \label{fig:gmrz}
\end{figure}

The number counts of galaxies in the PAUS mock and their redshift distribution were presented in \cite{Manzoni:2024}, who found good agreement with the PAUS observations. Manzoni et~al. also presented the median observed $g-r$ galaxy colour as a function of redshift. We present a slightly different view of this comparison between the mock and PAUS in Fig.~\ref{fig:gmrz}, in which we show the full distribution of observed $g-r$ colour in three narrow slices of redshift. This plot shows that before the introduction of photometric errors, the colour distribution in the mock is more sharply bimodal than for the real galaxies, a feature that has long been known in \texttt{GALFORM} (e.g. \citealt{gonzalez:2009}). After adding photometric errors, the distribution of colours in the mock matches the observed distribution more closely. Given the clear colour bimodality in the observations, and the fact that an observed property can be used to classify galaxies into two distinct populations, we apply the regression methods to red and blue galaxies separately in the next section. The red and blue populations are defined using the observed colour as a function of redshift given in eq. 2 of \cite{Manzoni:2024}. 

\begin{figure}
\includegraphics[trim= 0cm 0cm 0cm 0cm, width=0.99\columnwidth]{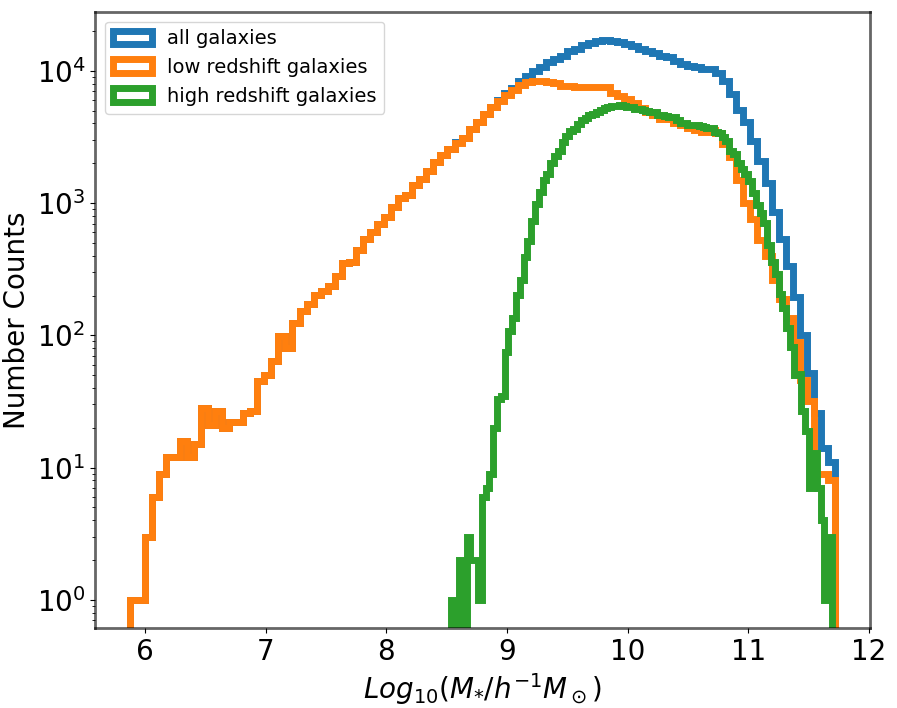}
    \caption{The distribution of stellar masses for galaxies in the PAUS mock with $i_{\rm AB} < 22.5$.
    The blue curve shows all galaxies, the orange curve galaxies with $z<0.55$ and the green curve $z>0.55$. The $y$-axis gives the raw number of galaxies in each bin, without adjustment for the bin width.
    }
    
\label{fig:mstars}
\end{figure}

The distribution of galaxy stellar masses in the full PAUS mock is shown in Fig.~\ref{fig:mstars}. The low mass tail extends to $10^{6} h^{-1} M_{\odot}$; these galaxies satisfy the sample selection in the apparent $i$-band magnitude, but will be extremely faint in absolute magnitude and exclusively at the lowest redshifts. The lightcone volume at low redshift is very small, hence the small numbers of these gaalxies. Later, we will see that training sets of galaxies for various regression methods are drawn at random from the distributions of stellar masses, so the mixture of stellar masses affects how well the connection between photometry and stellar mass can be learnt for different masses. 

As a final characterisation of the PAUS mock, we examine the correlation between stellar mass and galaxy brightness in different observed broad-band filters.

\begin{figure}
    \centering
    {\includegraphics[trim= 0cm 0cm 0cm 0cm, width=0.99\columnwidth]{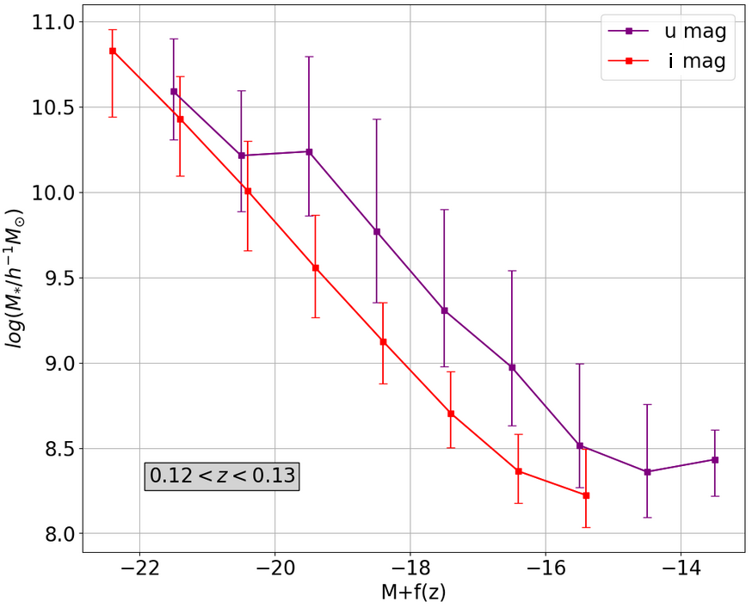}}
    {\includegraphics[trim= 0cm 0cm 0cm 0cm, width=0.99\columnwidth]{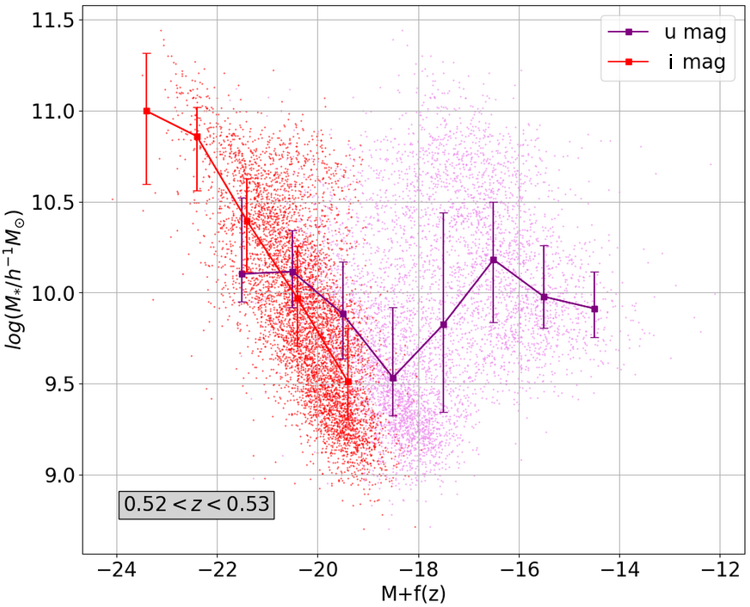}}
    {\includegraphics[trim= 0cm 0cm 0cm 0cm, width=0.99\columnwidth]{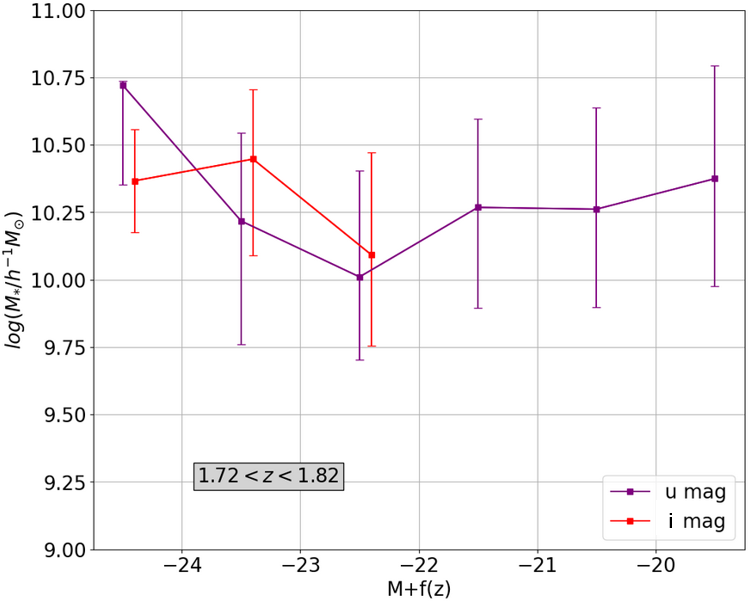}}
    \caption{
    The median stellar mass as a function of the minimal absolute magnitude (see text), plotted in narrow redshift slices: top - $0.12<z<0.13$, middle - $0.52<z<0.53$, bottom - $1.72<z<1.82$. The larger symbols show the median stellar mass and the bars shows the $25^{\rm th}$ to $75^{\rm th}$ percentile range; when there is a small number of galaxies in a bin, the 
    data points are not shown. Points are offset along the $x$-axis for clarity. The red points and lines are for the observed $i$-band and the purple points and lines for the observed $u$-band. In the middle panel we also show the individual galaxies, with purple dots showing the $u$-band and red dots the $i$-band, to help with the interpretation of the shapes of the median curves. 
    }

    \label{fig:naive}
\end{figure}

First, to include known redshift-dependent effects in the observed magnitudes, we take into account the dependence of the luminosity distance on redshift, $d_{\rm L}(z)$, assuming the cosmology adopted in the P-Millennium simulation. We use an approximate version of the absolute magnitude defined by:  


\begin{equation}
    M + f(z) = m - (5\log(d_{\rm L}(z)/h^{-1}{\rm Mpc}) + 25), 
\end{equation}
where $m$ is the observed apparent magnitude and $M+f(z)$ is an approximation to the absolute magnitude. We have purposefully left out the $k$-correction, which accounts for the change in the shape of the spectral energy distribution following the band shifting of the filter. The $k$-correction is galaxy-type dependent. We aim to effectively learn the $k$-correction from the galaxy brightness in the other pass bands, using machine learning or regression techniques. We call this magnitude, $M + f(z)$,  a minimally processed absolute magnitude because we have removed the known redshift effects, e.g. the luminosity distance, given the assumed cosmology, and have not attempted to correct for the model-dependent $k$-correction.

At low redshift, this minimally processed galaxy magnitude has a monotonic dependence on stellar mass as shown by the top panel in Fig.~\ref{fig:naive}. We could use the minimal absolute magnitude to read off the stellar mass, albeit with some scatter. Using the $i$-band, the 25-75 percentile range 
of stellar masses at a given magnitude is about a factor of two. For the $u$-band, the scatter is twice as big again, as expected because this band, at low redshift, is more sensitive to the recent star formation history than the $z$-band.  

With increasing redshift, the dependence of stellar mass on magnitude becomes more complicated, as shown by the middle and bottom panels of Fig.~\ref{fig:naive}. This is expected given the bandshifting of the observed magnitudes. This results in the filters in the observer frame sampling progressively shorter wavelengths in the rest frame spectral energy distribution with increasing redshift. The scatter in stellar mass increases with redshift for a given magnitude, and the relation is no longer monotonic. The middle panel of Fig.~\ref{fig:naive} shows individual galaxies as well as the binned median and interquartile range. This is to help understand the shape of the median curve of stellar mass versus magnitude, which is the most complicated in the middle panel of Fig.~\ref{fig:naive}. The relation at redshift $z=0.5$ in the $u$ band is bimodal, which results in the non-monotonic relation of the stellar mass on the absolute magnitude. These plots illustrate the challenge of estimating the stellar mass of a galaxy from photometry in a fixed filter set, particularly over a wide baseline in redshift. It is clear from these plots that a single band is insufficient to give an accurate estimate of the stellar mass over a range of redshift.

\section{Deep learning the stellar mass --  observables relation}

We first apply a so-called `black-box' machine learning method to learn the connection between the photometry and redshift of a galaxy and its stellar mass. Here, black box means that there is no specification of the functional form that the machine learning method should use to describe the relation between the observables (broad-band photometry and redshift) and stellar mass. Although there could be some modest variation in the performance of different machine learning methods for this task, we expect the drop in performance on moving to symbolic regression methods in the next section to be much greater. This is  because we are using more concise equations to describe the dependence of stellar mass on observables in symbolic regression, or equivalently much fewer parameters than are used by deep learning to capture the relation. Hence, the approach taken in this section will be considered as the benchmark, against which the various symbolic regression techniques presented in the next section will be compared.

We use the broad-band magnitudes and redshift from the PAUS mock as inputs to the machine learning method. We start with the unperturbed versions of these quantities. Since we expect them to have different star formation histories, red and blue galaxies are treated separately, and are defined using the observed $g-r$ colour given by eq.~2 from \cite{Manzoni:2024}. We divide each colour sample into training and testing datasets, using an 80:20.\footnote{Note we do not use a validation sample.}

We take a deep-learning approach using an artificial neural network (ANN) to describe the connection between the input properties (redshift and broadband photometry) and the output (stellar mass). The first layer of neurons 
is fed by the input properties. We discuss in the following the number of inputs used. The final layer of neurons contains the single output neuron, the stellar mass. In between these layers, we use 4 hidden layers, which contain 16, 32, 64 and 128 neurons. Through trial and error, we found that increasing the number of neurons in the hidden layers in this way helps avoid overfitting and allows more complexity to be captured between the inputs and outputs. We use a rectified linear unit (ReLU) activation function for each neuron\citep{Relu}.


\subsection{Performance metrics}

To assess the performance of a regression method, we define an overall `goodness of fit' metric, which is derived by applying the method to data it has not been trained on. 

We use the following metric 
\begin{equation}
   {\mathcal{M}} = \frac{1}{p}\sum_{1}^{p} |\log M_{\rm est} - \log M_{\rm true}|, 
\end{equation}
where $\log M$ is the logarithm of the stellar mass of the galaxy, $M_{\rm est}$ is the estimated mass, and $M_{\rm true}$ is the true galaxy stellar mass. The sum is over the $p$ galaxies in the testing sample. 
We have taken the modulus of the difference in the logarithm of the estimated and true stellar masses rather than the square to downweight the contribution from outliers.  


We also consider the bias in the mass estimate, $\log (M_{\rm est}/h^{-1}M_{\odot}) - \log (M_{\rm true}/h^{-1} M_{\odot})$ as a function of the true stellar mass, along with the 25-75 percentile range of the spread of the estimated masses (e.g. Fig. 4 and subsequent plots).  



\subsection{Choosing the number of input properties to use}

\begin{figure}
    \centering
    {\includegraphics[width=1.0\columnwidth]{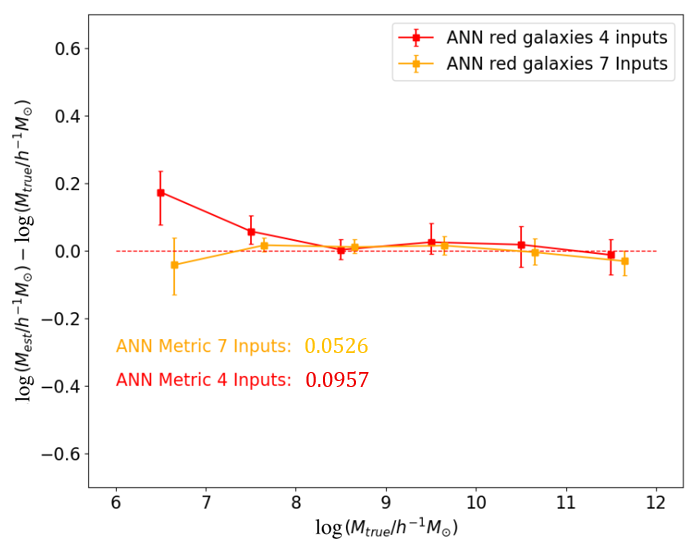}}
    {\includegraphics[width=1.0\columnwidth]{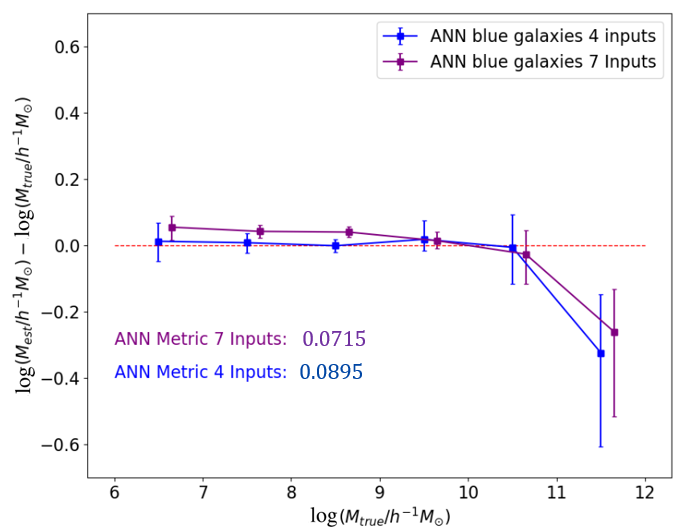}}
    \caption{The difference in the logarithm of the estimated stellar mass and that of the true stellar mass, plotted as a function of the log of the true stellar mass. Here we are comparing the performance of the ANN using 6 input parameters against using 4 input parameters, for both red and blue galaxies. There are 177416 red galaxies and 297996 blue galaxies. The bars show the 25-75 percentile spread of the predicted masses. 
    There is a modest increase in the scatter on the estimated stellar masses when restricting attention to four inputs. There is a tendancy for the most massive blue galaxies to have their stellar mass underestimated.  
    }
    \label{fig:num_inputs_test}
\end{figure}

We can draw the input properties from the four broad band magnitudes, corrected for known redshift effects as described above, the observed galaxy colors and the redshift. When we apply symbolic regression, we will sacrifice some level of accuracy by limiting the number of input properties used in the regression. The primary motivation for doing this is to restrict the complexity of the recovered function. By doing this, we give up some accuracy of reproduction for the sake of obtaining an expression that is easier to use. In any case, as we will see, the expressions we chose can be justified physically. Adding more terms to the expressions is likely to lead to overfitting. 

A further concern is co-linearity between properties, which refers to the properties not varying independently of one another. Ideally we want to avoid using properties that vary co-linearly with one another in our expressions, as this adds unnecessary complexity. Different functions of redshift (see below) are obviously co-linear. The broad-band magnitudes are co-linear to differing extents; the correlation is closer between bands if they are closer together in wavelength. Observed colours also have some co-linearity.   
 
We have considered two cases: (i)  six input properties: the $u$ and $i$, band magnitudes, the $g-r$ colour, redshift and two functions involving redshift, $z^{2}$ and $\log(1+z)$ and (ii) four input properties: the $u$ and $i$ broad band magnitudes, the $g-r$ colour and the redshift. In principle, we should not need to supply redshift and functions of redshift as inputs to the ANN. We nevertheless supplied the $(1+z)$ and $z^{2}$ terms as these mimic the shape of the $k$-correction as a function of redshift (see \citealt{Koonkor2025}). 

We contrast the performance of the ANN with six inputs against the one with four inputs in   Fig.~\ref{fig:num_inputs_test}. The top panel shows the results for red galaxies and the bottom panel those for blue galaxies. 

There is a slight reduction in performance (i.e., a slight \textit{ increase} in the metric) on decreasing the number of input properties used. The 25-75 percentile spread in the  predicted mass is slightly larger when using four inputs. However, in the interest of obtaining a simpler symbolic regression equation, we are ready to sacrifice the small gain in performance that results from using more input properties, and henceforth we will use four input properties.  The recovery of the stellar mass works to within $\approx 0.05$ dex for red galaxies. The performance of the ANN is similar for blue galaxies, for most stellar masses. However, the mass is significantly under estimated for the highest mass blue galaxies. The reason for this underestimate is not clear.                  

\subsection{Is it better to subdivide training samples by redshift?}

\begin{figure}
    {\includegraphics[width=1.00\columnwidth]{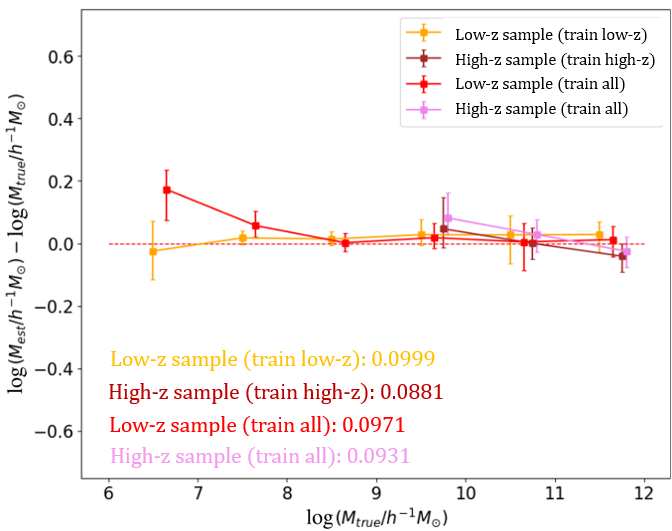}}
    {\includegraphics[width=1.00\columnwidth]{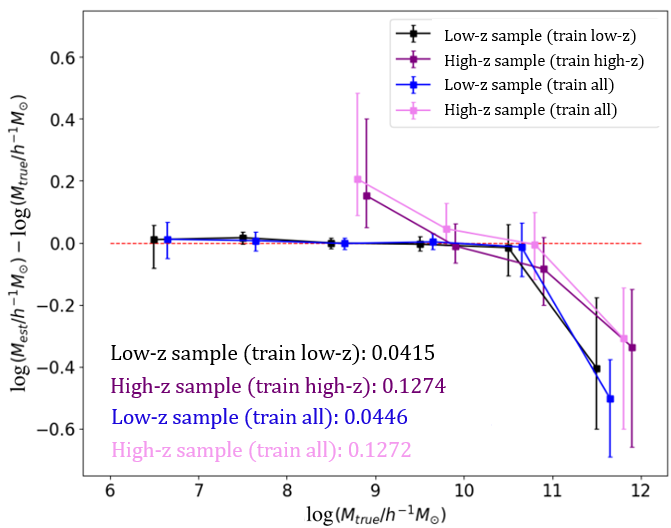}}
    \caption{The performance of the ANN for low and high redshift samples, when trained for (i) a single sample covering the whole redshift range and (ii) for low and high redshift samples separately. The top panel shows the results for red galaxies and the bottom panel those for blue galaxies. The metric value is labelled for each case and the bias in the estimated stellar mass is plotted as a function of the true stellar mass. The points show the median residual  and the bars the 25-75 percentile range. The key shows the colours used to denote each sample; these colours are also used to report the metric values. In the top panel, the red and purple lines show results for the ANN trained using all redshifts, when applied to all galaxies (red) and just to the high redshift sample (purple). Red should be compared with gold, which shows the low redshift galaxy predictions for an ANN trained using all galaxies. In the bottom panel the equivalent comparison is between blue and black points. }
    \label{fig:zsplit}
\end{figure}

The distribution of galaxy stellar masses changes with redshift in a magnitude-limited catalogue, as shown by Fig.~\ref{fig:mstars}. This raises the question of whether the performance of the 
ANN can be improved by splitting the sample in redshift as well as in colour, so that the ANN applied to high redshift galaxies is trained on a sample that better reflects the properties of these galaxies.

We test this idea by splitting the sample into high and low redshift subsamples using the median redshift, $z=0.55$, and training the ANN for each redshift subsample.  The results of this exercise are shown in Fig.~\ref{fig:zsplit}, which compares the metric, along with the bias and scatter for red (top panel) and blue (bottom panel) galaxies, so that we can compare the predictions from one ANN (trained and applied over all redshifts) and two ANNs (one trained and applied at low redshift and one trained and applied at high redshift). 

Each panel of Fig.~\ref{fig:zsplit} shows the results for one colour population i.e. red (top) and blue (bottom), and makes two comparisons. The first comparison is between the results for low redshift galaxies; in one case the ANN was trained using low redshift galaxies and in the other case the ANN was trained using galaxies of all redshifts. The second comparison is for high redshift galaxies, again considering the results obtained using an ANN trained on just the high redshift sample and those from an ANN trained using all galaxies. 

There is remarkably little difference in the recovered scatter. Even though the ANN trained at high redshift sees more examples of massive galaxies, as a fraction of the training population, there is only a modest change in the predicted scatter of those galaxies for the different training scenarios. Therefore, we will use an ANN trained over the full redshift range in the analysis below.

We note that there is a significant bias in the stellar mass estimated for high redshift galaxies. For the lowest stellar mass red galaxies in the high redshift sample, the stellar mass is overestimated by just under a factor of two; the mass of the most massive galaxies in this redshift range is underestimated by up to 10 per cent. Both of these factors are larger in the case of blue galaxies. The reason for the underperformance for massive blue galaxies is not immediately clear: it could be the case that the spectral features are weaker in these objects if a larger fraction of their stellar mass formed more recently.

\subsection{The impact of adding errors to the photometry and redshift}

\begin{figure}
    \centering
    \includegraphics[width=29em]{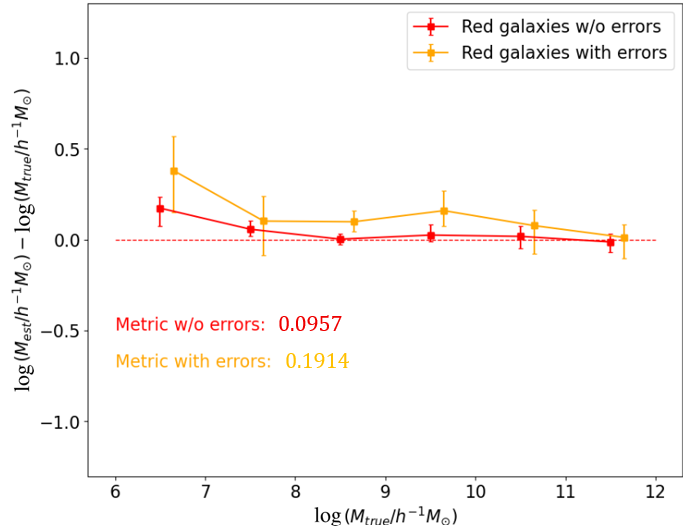}
    \includegraphics[width=29em]{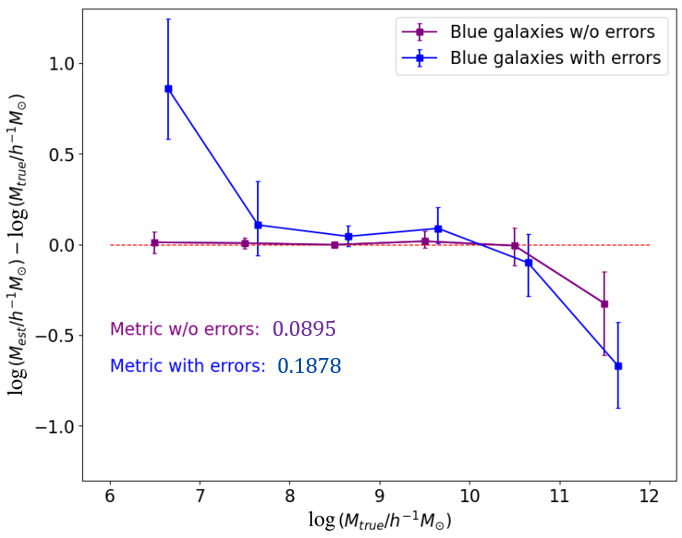}
    \caption{Comparing the performance of ANN for the ideal mock and the mock with photometry errors and photometric redshift errors added. The top panel shows the results for red galaxies and the bottom panel shows blue galaxies. 
    }
    \label{fig:ann-error}
\end{figure}

In real observations, errors in the measured photometry and in the estimated photometric redshift will affect the relation between observed galaxy properties and the stellar mass of the galaxy. Using the model for these errors set out in Section \ref{sec:err}, we now investigate how the ANN predictions are affected by these errors.

Fig.~\ref{fig:ann-error} compares the bias and scatter in the estimated stellar mass, along with the overall metric, for the cases without and with the observational errors. In each case the ANN is retrained. 
The ANN is sensitive to changes in the data given its intricate,  multiple layers and complexity. When we add the errors pertaining to photometry measurement and photometric redshift estimation, the performance of the ANN drops significantly, with the metric increasing by almost a factor of 2. Even with the observational errors included, the ANN is still able to give reasonable estimates of the stellar mass. We will see below that the symbolic regression models behave in a similar way to the ANN when observational errors are included, but with subtle differences.


\section{SYMBOLIC REGRESSION: DERIVING EXPLICIT STELLAR MASS RELATIONS}

\subsection{Motivation and approach}
While the ANN described in Section 3 provides accurate stellar mass estimates, it operates as a black box: the relationship between observables and stellar mass is encoded implicitly within thousands of network weights distributed across multiple layers. This lack of transparency makes it difficult to gain physical insight into which observables drive the mass estimation, and the model cannot be easily applied without the full network infrastructure.
To address these limitations, we apply symbolic regression methods to derive explicit mathematical expressions that relate stellar mass to observables. Symbolic regression seeks to find functional forms—combinations of mathematical operations and input variables—that best fit the data. Unlike traditional regression, which assumes a predetermined functional form (e.g., linear or polynomial), symbolic regression searches through a space of possible mathematical expressions to find optimal representations.
Our goal is not to achieve the maximum possible accuracy, but rather to find simple, physically interpretable expressions that perform reasonably well. This involves a deliberate trade-off: we sacrifice some predictive power in exchange for transparency and ease of use. The resulting equations can be evaluated instantly without specialized software and provide insight into how different observables contribute to stellar mass estimation.
Following the approach in Section 3, we continue to treat red and blue galaxies separately, using the same color-based classification from \cite{Manzoni:2024}. We use the same four input properties as in Section 3.2: the minimally processed absolute magnitudes in the $u$ and $i$ bands (Equation 3), the observed $(g-r)$ colour, and redshift. The training and test samples are identical to those used for the ANN.

\subsection{Symbolic regression methods}
We evaluate six different symbolic regression techniques, which vary in their complexity and the types of functional forms they can discover:
\begin{itemize}
\item[](i) Linear Regression \citep{Maulud:2020}: Models the stellar mass as a weighted sum of the input variables. Despite its simplicity, this approach can be effective when the relationship is approximately linear or when nonlinear effects are already captured through the choice of input features (e.g., using color as an input).
\item[](ii) Polynomial Regression \citep{Maulud:2020}: Extends linear regression by including polynomial terms (squares, cubes, cross-products) of the input variables. This can capture nonlinear relationships but at the cost of increased complexity and potential overfitting.
\item[](iii) Partial Least Squares (PLS) Regression \citep{GELADI:1986}: Particularly useful when input variables are correlated (collinear), as is the case with photometric bands. PLS reduces the inputs to a smaller set of uncorrelated components before performing regression, which can improve stability and generalization.
\item[](iv) Least Absolute Shrinkage and Selection Operator (LASSO) Regression \cite{Robert:1996}: Similar to linear regression but adds a penalty term that encourages sparse solutions—i.e., solutions that use fewer input variables. This helps identify which observables are most important for estimating stellar mass and can prevent overfitting.
\item[](v) Genetic Programming (GP) Regression \citep{Hongsup:2023}: Uses evolutionary algorithms to search through a large space of possible mathematical expressions. Starting with a population of random expressions, GP iteratively combines, mutates, and selects expressions based on their fit to the data, potentially discovering complex nonlinear relationships.
\item[](vi) Principal Component Analysis (PCA) \citep{Jollife:2016}: A dimensionality reduction technique that transforms the input variables into uncorrelated principal components ordered by the variance they explain. While not strictly a regression method, PCA-based approaches can simplify the input space before applying other regression techniques.
\end{itemize}

For all methods, we deliberately constrain the complexity of the resulting expressions. Rather than allowing arbitrary combinations of all possible terms, we restrict attention to expressions that include at most linear combinations of the four input properties. This constraint is motivated by several considerations: (i) simpler expressions are easier to interpret physically, (ii) overly complex expressions may fit noise rather than underlying relationships, and (iii) the resulting equations should be practical to use. We note that this approach differs from unconstrained symbolic regression, which might discover highly complex but potentially overfit expressions.

\subsection{Performance comparison}
\begin{figure}
    \centering
    \includegraphics[width=30em]{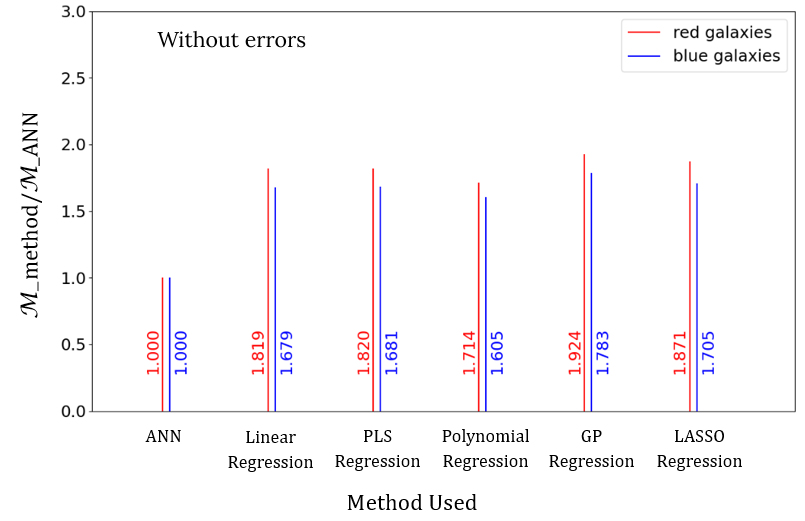}
    \includegraphics[width=30em]{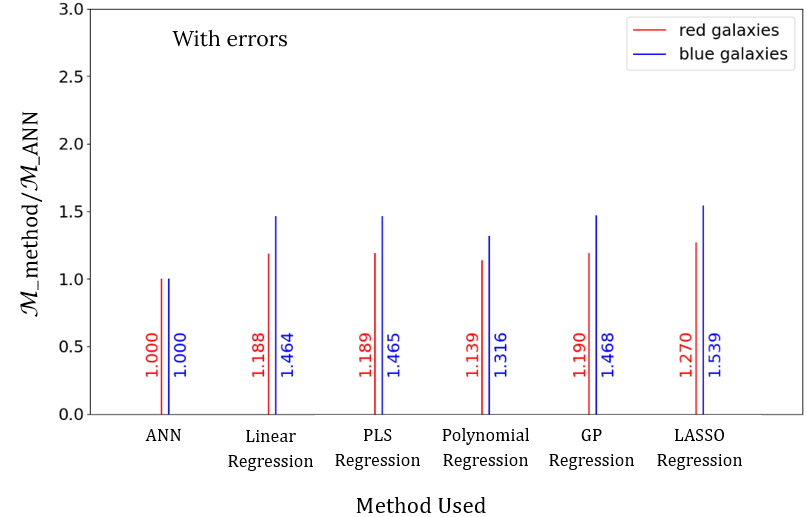}
    \caption{A comparison of the metric values obtained when estimating the stellar masses of red and blue galaxies using various symbolic regression techniques, as labelled, in units of the metric achieved using the benchmark ANN method. In the metric ratio is greater than one, then the method is performing worse than ANN; similar if we had a case for which the metric ratio was less than 1, then this method would be working better than the ANN. The top panel shows the comparison without any error, whereas the bottom panel includes both photometry and photometric redshift errors.  For example, a value of 1.0636 for the metric for red galaxies obtained using linear regression means that the metric in this case was just over 6 per cent larger than that obtained with the ANN for the same sample. }
    \label{fig:ann-symb-metric}
\end{figure}

We evaluate each method using the same metric defined in Equation 4 and compare the results against the performance of the ANN benchmark. Fig.~\ref{fig:ann-symb-metric} presents this comparison for both red and blue galaxies, with and without observational errors included in the photometry and photometric redshifts.
The top panel of Fig.~\ref{fig:ann-symb-metric} shows the performance when using the true (unperturbed) photometry and redshifts from the mock catalogue. All symbolic regression methods yield metrics approximately $1.5$–$2$ times larger than the ANN benchmark, indicating reduced accuracy compared to the deep learning approach. This is expected: the ANN can capture complex, nonlinear relationships through its multi-layer architecture and numerous parameters, while the constrained symbolic regression expressions are deliberately simplified. Among the methods tested, polynomial regression performs marginally better than the others, but at the cost of producing more complex expressions with numerous cross-terms. 

The performance comparison changes substantially when we include observational errors (bottom panel of Fig.~\ref{fig:ann-symb-metric}). Now the symbolic regression methods show metrics only 10–25 per cent larger than the ANN. This remarkable result demonstrates that symbolic regression is more robust to noisy inputs than deep learning. The complex, interconnected structure of the ANN means that input noise can propagate and amplify through the network layers, while the explicit functional forms in symbolic regression provide inherent smoothing that makes them less sensitive to input perturbations. Across all scenarios, linear regression emerges as the best-performing method when observational errors are included. While polynomial regression sometimes achieves marginally lower metrics in the error-free case, the improvement is modest and comes at significant cost in interpretability. We therefore adopt linear regression as our preferred approach. Tables 1 and 2 quantify the performance of each method more precisely, reporting both the mean metric value and its uncertainty estimated using bootstrap resampling (1000 trials, selecting galaxies randomly with replacement; \citealt{Norberg:2009}). Table 1 shows results for the ideal case without observational errors, while Table 2 includes photometry and photometric redshift errors. The errors on the metric values are remarkably small (typically $\sim 0.0003$–$0.0005$), reflecting the large sample sizes available for training and testing.

\begin{figure}
    \centering
    \includegraphics[width=30em]{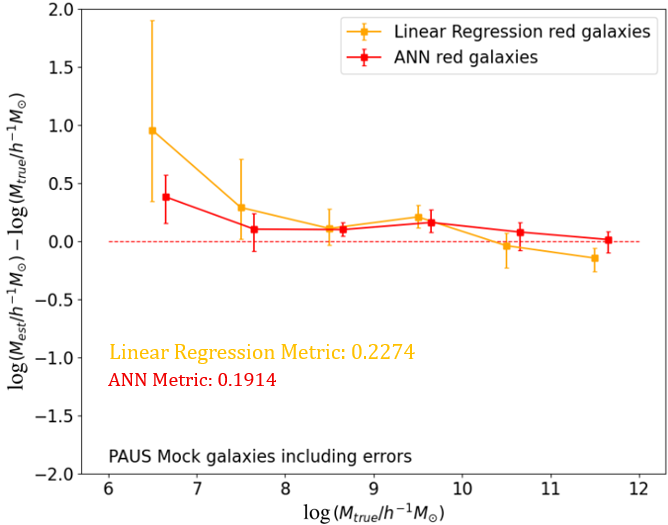}

    \vspace{0.8em}
    \includegraphics[width=30em]{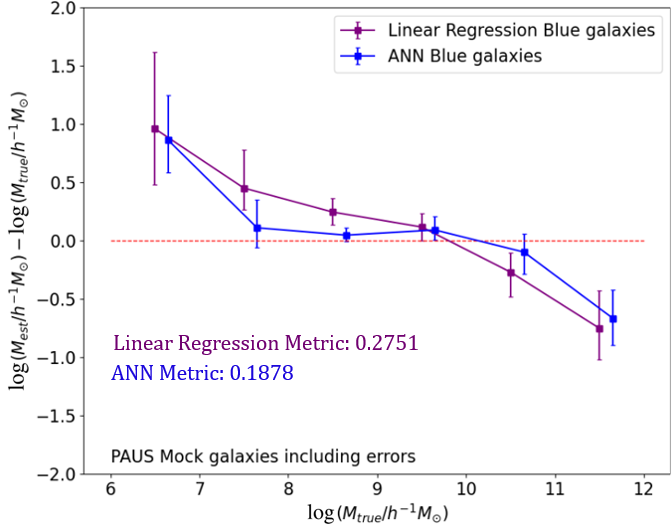}
     
    \caption{The performance of the best symbolic regression techniques compared with the predictions obtained using the benchmark ANN. The $x$-axis shows the true stellar mass and $y$-axis gives the bias in the recovered stellar mass, expressed in terms of the difference in the logarithms of the predicted and true stellar masses. Four input parameters (observables) are used as described in Section 4.2 and errors are included in the photometry and photometric redshifts of the model galaxies.}
    \label{fig:bestsymbregvsann}
\end{figure}

\subsection{The expressions for stellar mass}
Applying linear regression to the mock catalogue without observational errors yields the following expressions for estimating stellar mass:

\begin{equation}
\begin{aligned}
&{\rm red \,\, galaxies:} \\
&   \log(M_{*}/h^{-1}M_{\odot}) =  1.866 + 0.057 \times u_{\rm mag} +  \\
&  \quad \quad \quad 0.097 \times (g-r) - 0.457 \times i_{\rm mag} - 0.038 \times z  \\ \nonumber
&{\rm blue \,\, galaxies:}  \\
&    \log(M_{*}/h^{-1}M_{\odot}) = 1.725 + 0.305 \times u_{\rm mag}  \\ 
& \quad \quad \quad + 0.102 \times (g-r) - 0.674 \times i_{\rm mag} + 0.149 \times z, \nonumber
\end{aligned}
\end{equation}

When observational errors are included in the training data, we obtain:

\begin{equation}
\begin{aligned}
&{\rm red \,\, galaxies:} \\
&   \log(M_{*}/h^{-1}M_{\odot}) =  2.738 + 0.004 \times u_{\rm mag} +  \\
&  \quad \quad \quad 0.811 \times (g-r) - 0.334 \times i_{\rm mag} - 0.498 \times z  \\ \nonumber
& {\rm blue \,\, galaxies:}  \\
&    \log(M_{*}/h^{-1}M_{\odot}) = 3.524 + 0.209 \times u_{\rm mag}  \\ 
& \quad \quad \quad + 0.114 \times (g-r) - 0.493 \times i_{\rm mag} + 0.195 \times z. \nonumber 
\end{aligned}
\end{equation}

The functional form remains the same in both cases—a linear combination of the four input properties—but the numerical coefficients change substantially when errors are included. This reflects the method's adaptation to noisier data: the coefficients adjust to weight the inputs differently when measurement uncertainties are present.

\subsection{Physical interpretation}
These expressions reveal several interesting features about how observables relate to stellar mass.
The largest coefficients (in absolute value) are associated with the $i$-band magnitude for both red and blue galaxies. For red galaxies, the coefficient of $i_{\textrm{mag}}$ is 
 $−0.457$ ($−0.334$ with errors), indicating that this property is the primary mass indicator. For blue galaxies, both the $u_{\textrm{mag}}$ and $i_{\textrm{mag}}$ contribute significantly, with coefficients of $0.305$ and $−0.674$ respectively ($0.209$ and $−0.493$ with errors). 

The importance of the $i$-band reflects the fact that at the typical redshift of the PAUS sample ($z \sim 0.5$), this band probes rest-frame optical wavelengths that trace the bulk stellar population. The additional contribution from the $u$-band for blue galaxies captures the recent star formation that produces UV emission, which is important for estimating the total stellar mass in galaxies with ongoing star formation.

Both expressions show positive coefficients for the observed $g-r$ colour ($0.097$ for red galaxies, $0.102$ for blue galaxies in the error-free case). Redder galaxies have larger $g-r$ values and, all else being equal, higher estimated masses. This makes physical sense: red colors typically indicate older stellar populations with higher mass-to-light ratios. Interestingly, when observational errors are included, the coefficient for red galaxies increases substantially to $0.811$, suggesting that colour becomes a more important discriminant when individual magnitudes are noisier.

The expressions show that stellar mass estimates decrease weakly with redshift for red galaxies (coefficient of $−0.038$, or $−0.498$ with errors) but increase with redshift for blue galaxies (coefficient of $+0.149$, or $+0.195$ with errors). These terms effectively capture some of the $k$-correction and evolutionary effects. As redshift increases, the observed bands shift to shorter rest-frame wavelengths. For red galaxies, which have little UV emission, this bandshifting means the observed magnitudes become fainter relative to the stellar mass, requiring a negative correction. For blue galaxies, with stronger UV emission, the effect works in the opposite direction.

We can understand these expressions as implicitly learning an effective $k$-correction from the mock data. Rather than computing the $k$-correction explicitly (which requires assumptions about galaxy spectral types), the symbolic regression discovers how to correct for bandshifting effects using the available observables. The fact that linear expressions can capture these effects reasonably well (as we will see in Section 4.6) suggests that, for the magnitude-limited sample considered here, the dominant systematics are indeed approximately linear in the chosen observables.

Finally, it is instructive to compare our empirically derived expressions with physically motivated stellar mass estimators. \cite{Robotham2020} proposed a simple mass estimator for galaxies based on the luminosity in a rest-frame optical band plus a correction for colour

\begin{equation}
\log \left( M_{*} /M_{\mathrm{\odot}} \right) 
\sim -0.4 \times {\mathrm{M}} + 
\beta  \times 
{(\mathrm{m_{1} - m_{2}})} + 
{\textrm{constant}}, 
\end{equation}
where the coefficient on magnitude (−0.4) comes from converting magnitude to luminosity, $M$ is an absolute magnitude, $\mathrm{m}_{1} - \mathrm{m}_{2}$ is a colour and $\beta$ depends on the choice of color, which in turn depends on the redshift of the galaxy. Our expressions have the same functional form, with magnitude terms, a color term, and redshift corrections. The coefficients we derive are consistent with this physical expectation.

\begin{table}
\centering
\begin{tabular}{lcc} 
\hline
ML method & mean & scatter \\
 \hline
 Artificial neural & 0.0957 & 0.00031 \\ 
 networks & 0.0895 & 0.00022  \\ 
 \hline
 Linear & 0.1741 & 0.00041 \\ 
 regression &0.1502 & 0.00029  \\
 \hline
 PLS & 0.1742 & 0.00044 \\ 
 regression &0.1504 & 0.00030  \\
 \hline
 Polynomial & 0.1640 & 0.00043 \\ 
 regression & 0.1436 & 0.00028 \\
 \hline
 LASSO & 0.1757 & 0.00042 \\ 
 regression & 0.1526 &  0.00029  \\
 \hline
 GP & 0.1842 & 0.00045 \\ 
 regression & 0.1595 & 0.00029  \\
 \hline
\end{tabular}
\caption{The metric values (column 2) and scatter (coulumn) for the mock without applying photometry and redshift errors. The first column gives the method applied to estimate the stellar mass. For method there are two rows; the first gives the values for red galaxies and the second row those for blue galaxies.}
\end{table}

\begin{table}
\centering
\begin{tabular}{lcc} 
 \hline

ML method & Mean & Scatter\\
 \hline
 Artificial Neural & 0.1886 & 0.00040 \\ 
 Networks & 0.1828 & 0.00037  \\ 
 \hline
 Linear & 0.2241 & 0.00050 \\ 
 Regression & 0.2677 & 0.00051  \\
 \hline
 PLS  & 0.2242 & 0.00050 \\ 
 regression & 0.2668 & 0.00053  \\
 \hline
 Polynomial & 0.2150 & 0.00049 \\ 
 Regression & 0.2407 & 0.00045  \\
 \hline
 LASSO  & 0.2243 & 0.00052 \\ 
 regression &0.2685 &  0.00053  \\
 \hline
 GP  & 0.2397 & 0.00061 \\ 
 Regression & 0.2814 & 0.00067  \\
 \hline
\end{tabular}
\caption{The metric values (column 2) and scatter (coulumn) for the mock including  photometry and redshift errors. The first column gives the method applied to estimate the stellar mass. For method there are two rows; the first gives the values for red galaxies and the second row those for blue galaxies.}
\end{table}

\subsection{Performance on mock data}
Fig.~\ref{fig:bestsymbregvsann} compares the performance of the linear regression expressions against the ANN benchmark, using the mock sample with observational errors included. The figure shows the bias in recovered stellar mass (difference between estimated and true log stellar mass) as a function of true stellar mass, separately for red and blue galaxies. For red galaxies (top panel), the linear regression expressions perform remarkably well across most of the stellar mass range. The median bias is within $\pm 0.05$ dex for masses $10^{9} <  M_{*}/(h^{-1} M _{\mathrm{\odot}})< 10^{11.5}$, and the 25–75 percentile scatter (shown by the error bars) is comparable to that achieved by the ANN. At the lowest masses ($M_{*}<10^{9} h^{-1} M_{\mathrm{\odot}})$, there is a tendency for both methods to overestimate the stellar mass by up to 0.1 dex. This is likely to reflect the magnitude limit of the survey: low-mass galaxies near the detection threshold have larger photometric uncertainties, and the fitting process tends to scatter them upward in mass.

For blue galaxies in the bottom panel of Fig.~\ref{fig:bestsymbregvsann}, the performance is similar at intermediate masses but shows larger systematic biases at the extremes. Low-mass blue galaxies show the same upward bias seen for red galaxies. More problematically, the most massive blue galaxies ($M_{*} > 10^{11} h^{-1} M_{\mathrm{\odot}}$) have their masses systematically underestimated by 0.1–0.2 dex. This is a feature also present in the ANN predictions (Section 3.2, Figure 4) and probably reflects the challenge of estimating masses for rare objects that may have unusual star formation histories or spectral properties not well represented in the training data.

The similarity in performance between linear regression and the ANN is particularly striking given the vast difference in model complexity. The ANN uses thousands of parameters across multiple layers, while the linear expressions involve just five coefficients. This demonstrates that, for this problem, the relationship between observables and stellar mass can be captured adequately by simple functional forms—at least when considering the ensemble properties of galaxies rather than attempting to achieve maximum accuracy for individual objects.

\subsection{Summary}
We have shown that simple, explicit linear expressions can estimate galaxy stellar masses from photometry and redshift with an accuracy approaching that of complex deep learning models, particularly when observational errors are present. 
The expressions we derive: (i) are physically interpretable, with coefficients that reflect expected dependencies on luminosity, color, and redshift, (ii) can be evaluated instantly without specialized software and (iii) are robust to observational noise. They 
perform similarly to expressions motivated by physical models \citep{Robotham2020}.

The main limitation is reduced accuracy for the most massive blue galaxies and for low-mass galaxies near the survey limit. Nevertheless, for the bulk of the galaxy population, these expressions provide a fast, transparent alternative to both traditional SED fitting and black-box machine learning approaches.

\section{Comparing stellar mass predictions: symbolic regression vs SED-fitting}

We now evaluate the performance of the machine learning-based stellar mass predictions on real observational data, We make use of the expressions found to predict the stellar masses for the mock catalogue, including errors in photometry and photometric redshift, and apply these to the PAUS observations. The input features required for these predictions--the photometric redshift $z_\texttt{obs}$, the observed $u$- and $i$-band magnitudes, and the observed $(g-r)$ colour--were obtained from the PAUS production photometry.

To assess the validity of the machine learning-based predictions, the resulting stellar masses were compared to those derived from spectral energy distribution (SED) fitting to the PAUS photometry using the \texttt{CIGALE} code by \cite{Csizi2024}. The \texttt{CIGALE} stellar masses are provided as part of the official PAUS production data. The comparison was performed separately for red and blue galaxies in the W1 and W3 fields.

\begin{figure}
\centering
    {\includegraphics[trim=0.5cm 0cm 0.5cm 2cm, clip=true,width=0.54\textwidth]{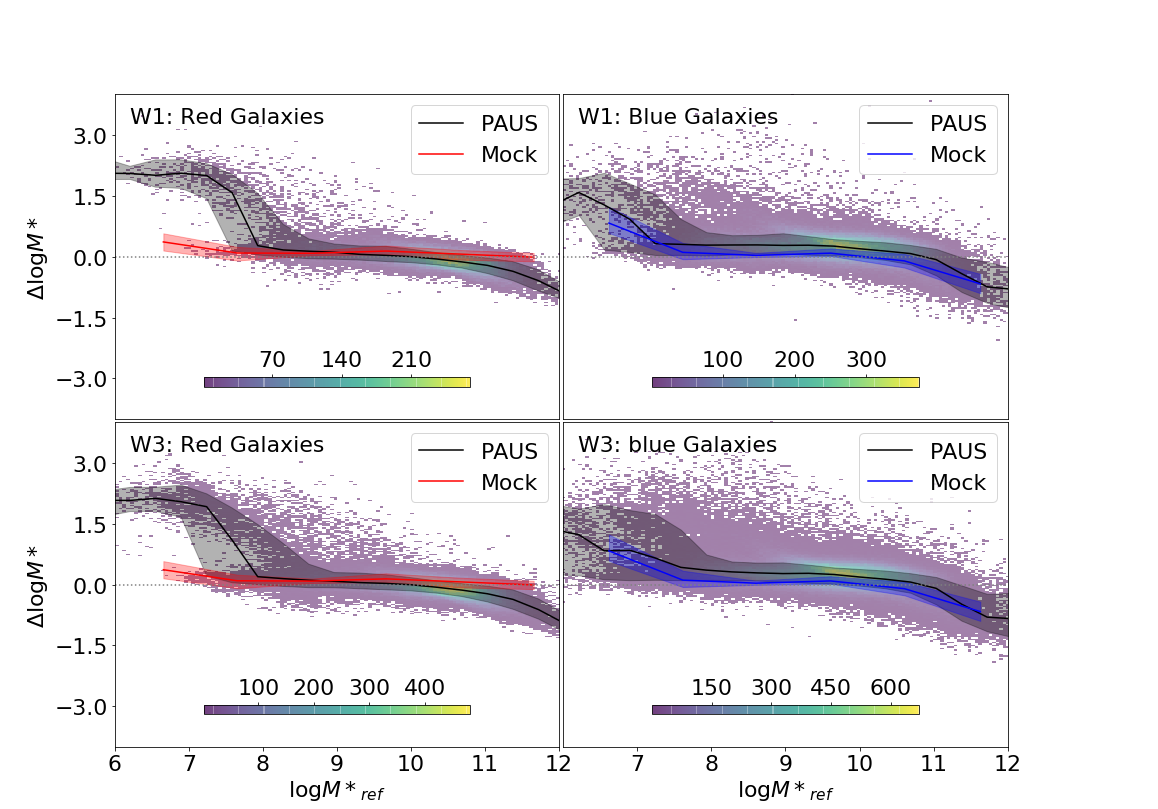}}%
 \caption{Comparison between stellar masses predicted using the symbolic regression expressions and those derived from SED fitting using the \texttt{CIGALE} code, for PAUS galaxies in the W1 (top row) and W3 (bottom row) fields. Each panel shows the difference in two values of the log stellar mass as a function of the \texttt{CIGALE}-derived reference mass (for the PAUS data -- here the $y$-axis shows the difference in the logarithm of the stellar mass from the regression expression and the log of the stellar mass from \texttt{CIGALE}) or the "true" mock stellar masses (for the \texttt{GALFORM} mock galaxies - here the $y$-axis shows the difference in the logarithm of the stellar mass from regression and that of the true stellar mass). The left and right columns correspond to red and blue galaxies, respectively. Solid lines show the running median trends for PAUS (black) and mock predictions (red or blue), with shaded regions indicating the 25th-75th percentile range. The background shading shows the density of galaxies in the PAUS sample in this plane. The overall agreement is generally good, with the residuals typically below 0.1 dex across most of the stellar mass range.}%
    \label{fig:CIGALEvsSymb_StellarMasses}
\end{figure} 

Fig.~\ref{fig:CIGALEvsSymb_StellarMasses} shows the difference between the stellar mass predicted using the symbolic regression expressions and the \texttt{CIGALE} stellar mass from SED fitting, plotted as a function of a reference stellar mass $\log M_{\ast, \textrm{rest}}$, for red and blue galaxies in both the W1 and W3 fields. For the case of PAUS, the reference stellar mass is taken to be the \texttt{CIGALE} stellar mass, whereas for mock galaxies the reference stellar mass is taken to be the galaxy stellar mass output by the simulation. The 2D density histograms show the distribution of individual galaxies from the PAUS data, while the black solid lines indicate the running median trend, with the shaded grey region showing the interquartile (25th-75th percentile) range. The red and blue solid lines represent the mock results for red and blue galaxies, including errors, as discussed in Fig.~\ref{fig:ann-error}. The general agreement between the two methods is good in the mass range $10^8 < M_\ast/h^{-1}M_\odot < 10^{11.5}$, with systematic offsets generally below 0.1 dex. The discrepancies become more significant at the low-mass end, where we expect the galaxies to be fainter and more prone to photometry errors, and at the high-mass end, where sample statistics are limited and the assumption about dust extinction made in \texttt{CIGALE} could differ from that calculated in \texttt{GALFORM}. 

These results show that the symbolic regression model provides a fast and reasonably accurate approximation to SED-based stellar mass estimates, with significant advantages in computational speed.

\begin{figure}
\centering
    {\includegraphics[width=0.48\textwidth]{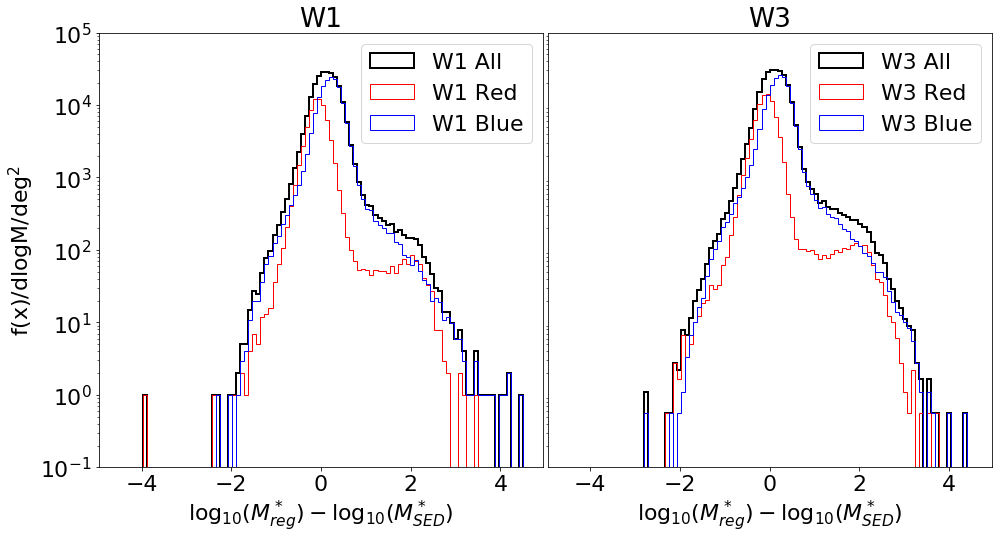}}%
 \caption{The distribution of the differences between the stellar masses from the observational data from the PAUS W1 field (left panel) and W3 field (right panel) estimated using the regression ($M^*_\textrm{reg}$) and \texttt{CIGALE} SED fitting code ($M^*_\textrm{SED}$). In each panel, the black line shows the distribution of all galaxies in each sample, while the red and blue lines represent the observed red and blue populations, respectively. The distributions are normalised with the covered areas and d$\log$M bin size (see \citealt{Koonkor2025} for the area covered by each field).}%
    \label{fig:dlogM_pdf}
\end{figure}

How closely should we expect the two estimates of the stellar masses to agree? To answer this question, we can look at the distribution of the logarithms of the mass difference for each galaxy when the two methods are applied: $\log M^{*}_{\textrm{reg}} - \log M^{*}_{\textrm{CIGALE}}$. This distribution is plotted in Fig.~\ref{fig:dlogM_pdf}. The mean of this distribution is 0.1348 dex for red and blue galaxies together.  
The \texttt{CIGALE} and \texttt{GALFORM} models both assume stellar initial mass function (IMF) forms compatible with solar neighbourhood determinations, the \cite{Chabrier2003} and \cite{Kennicutt1998} forms, respectively. Without applying any corrections, we would expect this difference in the IMF to give (using Table B1 from \citealt{Lacey:2016}): $\log M ^{*}_{\textrm{reg}} - \log M^{*}_{\textrm{CIGALE}} = \log{0.81} = -0.09$.  So the difference in the solar neighbourhood IMFs assumed in the two approaches would explain a smaller magnitude of offset than we find, and with the opposite sign.

 Fig.~\ref{fig:CIGALEvsSymb_StellarMasses} shows that applying this factor alone would bring the stellar mass estimates for red galaxies into better agreement. Applying the same correction to blue galaxies actually makes the correspondence between the two mass estimates slightly worse. The other differences in model assumptions, namely dust attenuation, treatment of stellar metallicity and the assumption of the form of star formation history used in \texttt{CIGALE} are harder to pick apart and contribute to the remaining biases and offset, and some component of the scatter (for a discussion of some of these points, see for example \citealt{Mitchell2013}). Nevertheless, the observation that adjusting for the different choices of the IMF brings the stellar mass estimates of red galaxies into better agreement and makes those of the blue galaxies worse suggests that the dust attenuation may be stronger in \texttt{CIGALE} than in \texttt{GALFORM}.
 It is interesting to note that for low mass red galaxies, there is a step in the accuracy of the predictions from \texttt{CIGALE} with bimodality in the distribution of mass estimates.

\begin{figure}
\centering
    {\includegraphics[width=0.48\textwidth]{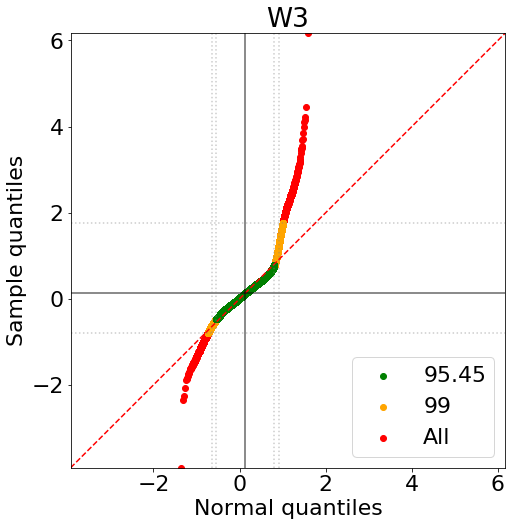}}%
 \caption{QQ plot comparing the quantiles for the ordered distribution of the logarithm of the mass differences for symbolic regression and SED fitting, for galaxies in the W3 field, which is the larger of the two fields, thereby allowing us to probe further into the tails of the distribution. If this distribution was Gaussian, all of the points would lie along the diagonal. The central 95.4 per cent of the distribution, coloured green, follows the Gaussian expectation. The extension to the central 99 per cent is shown in gold and the remaining points are shown in red.}%
    \label{fig:QQ}
\end{figure} 
\begin{figure}
    \centering
    \includegraphics[width=0.43\textwidth]{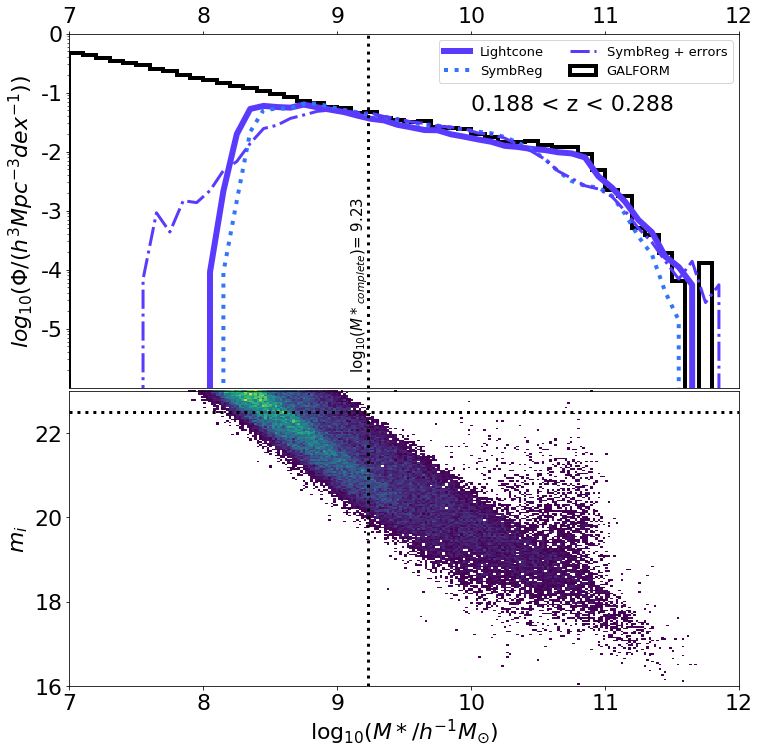}
    \includegraphics[width=0.43\textwidth]{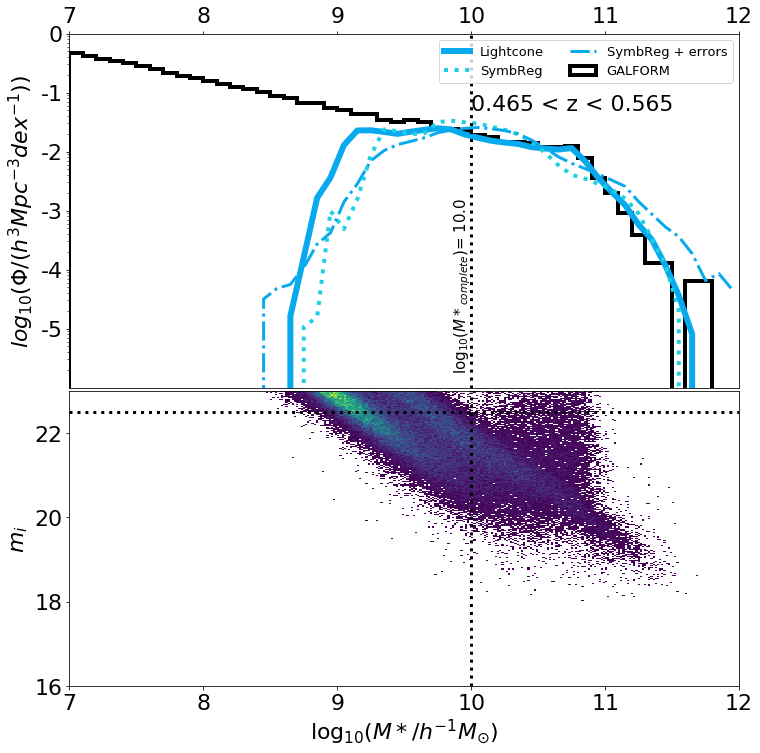}
    \caption{The stellar mass function at $z=0.2$ (top panels) and $z=0.55$ (bottom panels). For each redshift, the upper panel shows the stellar mass function and the lower panel shows the distribution of model galaxies plotted in the stellar-mass - $i$-band apparent magnitude plane, with brighter shading indicating a higher number of galaxies per pixel. In the stellar mass function plots, the \texttt{GALFORM} stellar mass function, using true stellar masses and without any observational selection, is shown by the black histogram. The stellar mass function measured from mock catalogue galaxies with $i_{\rm AB} < 22.5$ is shown by the blue solid line (cyan; bottom). This departure from the black histogram at low masses is due to observational selection, roughly at the mass indicated by the vertical dotted line in the lower panels. The dotted lines show the stellar mass function estimated from the mock using symbolic regression, without observational errors. The dot-dashed line shows the stellar mass function with photometry and photometric errors included.}
    \label{fig:mstarsz}
\end{figure}

\begin{figure}
    \centering
    \includegraphics[trim=0.cm 0cm 0cm 0cm, clip=true, width=8.5cm]{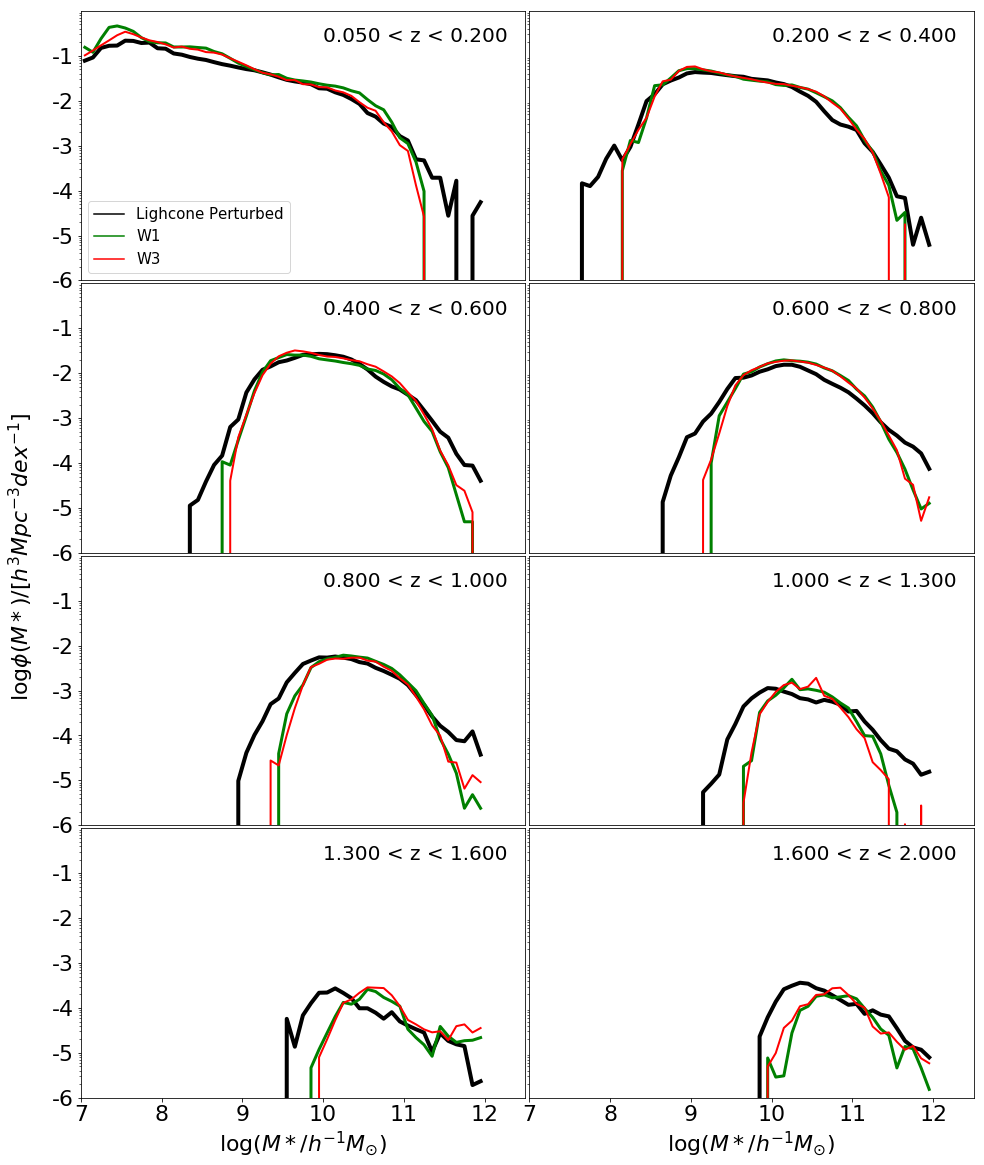}
    \caption{The stellar mass functions estimated using symbolic regression for different redshift bins as labelled in each panel. The black curves show the model predictions made using the lightcone galaxies, including photometry errors and photometric redshift errors. The green and red curves show the estimates from the PAUS W1 and W3 fields, respectively, using the same symboolic regression equations that were applied to the lightcone gaalxies. We show the W1 and W3 fields separately,  to show the impact of sample variance.}
    \label{fig:mstarszPAUS}
\end{figure}

We note that the distribution of the difference in the logarithm of the mass plotted in Fig.~\ref{fig:dlogM_pdf} have non-Gaussian tails. To better quantify this, we can plot a quartile-quartile plot (or Q-Q plot), which compares the distribution to a Gaussian. If the mass difference distribution was a Gaussian, the Q-Q plot would be a diagonal line. Fig.~\ref{fig:QQ} shows that this is indeed the case for the central 95 per cent of the distribution of mass differences, but with strong departures for the tails.

The \texttt{CIGALE} stellar mass estimates are supplied with an estimate of the associated error. The symbolic regression methods do not provide an error estimate. We can examine the mean and variance of the mass estimate distributions and the distribution of mass differences. 
The variance of the mass estimates is much larger than the variance of the mass difference. We also computed the covariance between the mass estimates and found this to be comparable to the variances. Hence, the \texttt{CIGALE} and symbolic regression mass estimates are correlated, which explains why the variance of the mass differences is smaller than the variance of either mass estimate. Finally, we note that the variance of the symbolic mass estimates is slightly smaller than that of the distribution of \texttt{CIGALE} estimates. 

\section{The stellar mass function}

We now use the expressions found for estimating the stellar mass of red and blue galaxies to estimate a basic property of the galaxy population, the stellar mass function. 
The galaxy stellar mass function gives the number density of galaxies as a function of their stellar mass and can be used to test galaxy formation models. To compensate for observational sampling effects, a weight is estimated for each galaxy using the $1/V_{\textrm{max}}$ method of \cite{Schmidt:1968}, as described in detail in \cite{Koonkor2025}. This allows us to correct the observed number of galaxies to obtain an estimate of the underlying number density of galaxies. 

We first consider the stellar mass function at selected redshifts from the range sampled by the PAUS mock catalogue. The $i$-band selection changes the form of the recovered stellar mass function, compared to the underlying mass function in the model.  
We are also interested in how the errors and biases introduced by the symbolic regression methods change the form of the recovered stellar mass function compared to the prediction using the true stellar masses in the model. The incorporation of errors in the photometry and photometric redshifts of the model galaxies can further distort the recovered stellar mass function. 

We show various model predictions in Fig.~\ref{fig:mstarsz}. The upper panels show the results in a narrow redshift bin centered on $z=0.2$ and the lower panels show the results at $z \sim 0.5$, which is close to the peak of the redshift distribution for a magnitude limit of $i_{\rm AB} = 22.5$. In both cases, the upper panel shows the stellar mass function. The black line histogram shows the true \texttt{GALFORM} stellar mass function at the central redshift in each bin as in the simulation box. This is the stellar mass function without any observational selection. The lower panel shows the distribution of galaxies from the lightcone plotted as a function of stellar mass and $i$-band apparent magnitude. This panel helps us to understand why the stellar mass functions recovered from the mock catalogue differ from the one estimated from the simulation box. The horizontal line in the lower panels shows the cut $i_{\rm AB}=22.5$. The vertical line has been placed by eye and indicates the lowest stellar mass for which all of the predicted distribution of galaxies is shown, given the $i$-band limit. Below this mass the recovery of the stellar mass function is incomplete. The locus of galaxies in the stellar mass - apparent $i$-band magnitude changes as we replace the true stellar mass by the mass estimated using symbolic regression. 

The upper panels of Fig.~\ref{fig:mstarsz} shows that the lightcone, using the true stellar masses, recovers the underlying mass function (as shown by the black histogram, plotted from the simulation volume at the central redshift in each panel, without any selection effects) down to stellar mass of $\approx 10^{9} h^{-1} M_{\odot}$. When plotting the stellar mass function using the masses returned by the symbolic regression equations, there are differences from the lightcone mass function at both low and high stellar masses. These differences become larger when the photometry and photometric redshift errors are included.  These discrepancies also increase with redshift. At $z=0.5$ the recovered mass function becomes incomplete below $\sim 10^{9.7} h^{-1} M_{\odot}$. At high masses, $10^{11.5} h^{-1} M_{\odot}$, the stellar mass function derived from symbolic regression, including observational errors, is a factor of $\sim 3$ higher than the true mass function.  At both redshifts, the masses returned by symbolic regression alter the shape of the mass function around the knee or break. This drop in performance can be traced to the band shifting of the observed magnitudes and colours, which means that a narrrower part (and shorter wavelengths) of the galaxy spectral energy distribution is sampled with increasing redshift.

Nevertheless, despite the scatter and bias obtained when estimating the mass of a single galaxy, it is possible to make a reasonably accurate recovery of the stellar mass function using symbolic regression. The mock allows us to assess where the recovery breaks down and by how much the recovered stellar mass function is in error. 

Finally, we show the estimate of the stellar mass function using the PAUS data in Fig.~\ref{fig:mstarszPAUS}. The results for the W1 (green) and W3 (red) are shown separately, to allow us to see the impact of field-to-field variations on the form of the mass function. These estimates are compared to the lightcone mock (black curves), which includes errors in the photometry and the  estimated photometric redshifts. As Koonkor et~al. found to be the case with the $i$-band luminosity, the agreement with the model is best at low and intermiediate redshifts. Discrepancies appear at the low and high stellar mass ends with increasing redshift.

\section{Summary and Conclusions}

We have presented a new approach to estimating galaxy stellar masses from photometric observations that combines the accuracy of modern machine learning with the interpretability of explicit mathematical formulae. Rather than relying on traditional forward modelling through spectral energy distribution (SED) fitting—which requires assumptions about star formation histories, stellar metallicity evolution, and dust attenuation—we trained regression algorithms using mock catalogues generated by the \texttt{GALFORM} semi-analytical galaxy formation model. This approach is similar in spirit to recent studies using hydrodynamical simulations \citep{Davidzon2019} or semi-analytical models \citep{Simet2021}, but we specifically focus on deriving simple, transparent expressions that can be applied rapidly to large datasets.

We first established a performance benchmark using an artificial neural network (ANN) with four hidden layers to learn the relationship between stellar mass and observables (minimally processed $u$- and $i$-band magnitudes, observed $g-r$ colour, and redshift). Following \cite{Manzoni:2024}, we treated red and blue galaxies separately based on their observed colours. The ANN achieves accurate stellar mass recovery for most of the galaxy population, with typical uncertainties of $\approx 0.05$ dex for red galaxies. However, the most massive blue galaxies ($M_* > 10^{11} h^{-1}M_\odot$) show systematic underestimates approaching a factor of two. The origin of this bias remains unclear but may reflect weaker spectral features in massive galaxies with recent star formation.

To gain physical insight and computational efficiency, we applied six symbolic regression techniques to derive explicit mathematical expressions for stellar mass. By deliberately constraining the complexity of these expressions to linear combinations of the input observables, we achieved a balance between accuracy and interpretability. Linear regression emerged as the optimal method, particularly when realistic observational errors were included. The resulting expressions take the form $\log(M_*/h^{-1}M_\odot) = c_0 + c_1 u_{\rm mag} + c_2 (g-r) + c_3 i_{\rm mag} + c_4 z$, with different coefficients for red and blue galaxies. These expressions are physically motivated: the dominant terms correspond to luminosity indicators ($i$-band for red galaxies; $u$- and $i$-bands for blue galaxies), the colour term captures variations in mass-to-light ratio, and the redshift term provides an effective $k$-correction. The functional form closely resembles physically motivated mass estimators proposed by \citet{Robotham2020}.

When observational errors in photometry and photometric redshifts are included—the realistic scenario for survey data—the symbolic regression performance degrades only modestly (metrics increase by 10--25 per cent) compared to the ANN benchmark. Remarkably, the ANN performance degrades more substantially than symbolic regression when noise is added to the inputs. This demonstrates that simple explicit formulae are more robust to observational uncertainties than complex neural networks, whose interconnected structure allows noise to propagate and amplify through multiple layers.

We validated our approach by comparing stellar mass estimates from symbolic regression with those derived from traditional SED fitting using the \texttt{CIGALE} code \citep{Csizi2024} applied to PAUS photometry. For galaxies with $M_* > 10^8 M_\odot$, the two methods show good agreement over most of the stellar mass range, with typical systematic offsets below 0.1~dex and a mean offset of 0.13~dex. The offset between the two mass estimates can be partially explained by the different stellar initial mass functions (IMFs) assumed: Chabrier in \texttt{CIGALE} versus Kennicutt in \texttt{GALFORM}, which would contribute $-0.09$~dex to the mass difference. Applying this IMF correction improves agreement for red galaxies but worsens it for blue galaxies, suggesting additional systematic differences in dust attenuation treatment between the two approaches. This is consistent with expectations, as dust attenuation is one of the largest sources of uncertainty in SED fitting \citep{Mitchell2013}.

At low masses ($M_* < 10^8 M_\odot$), we observe bimodality in the mass residuals: approximately one-third of galaxies show small offsets consistent with the behaviour at higher masses, while the remainder show larger discrepancies with symbolic regression systematically predicting higher masses. This bimodality likely reflects the combined effects of larger photometric uncertainties for faint galaxies near the survey limit and differences in how the methods handle dust extinction and star formation histories in low-mass systems. The distribution of mass differences exhibits non-Gaussian tails beyond the central 95\%, with the core of the distribution following Gaussian statistics (Figure~11). This has important implications for uncertainty quantification: simple Gaussian error propagation will underestimate the true uncertainties, particularly in the tails of the distribution. Importantly, the symbolic regression approach achieves this level of accuracy with negligible computational cost compared to SED fitting. Once trained, our expressions can be evaluated instantaneously for millions of galaxies, making them particularly attractive for upcoming large surveys such as LSST \citep{Ivezic2019ApJ} and Euclid \citep{Enia2024}, where SED fitting for billions of galaxies would be computationally prohibitive.

We tested whether our symbolic regression expressions could recover the stellar mass function—a fundamental diagnostic of galaxy evolution. The main observational selection (the $i$-band apparent magnitude limit) introduces the expected turnover at low masses, with the completeness limit increasing with redshift. The symbolic regression mass estimates introduce additional systematic effects at both low and high masses. At low masses, incompleteness from the magnitude limit combines with the tendency to scatter low-mass galaxies upward in estimated mass due to photometric uncertainties. At high masses ($M_* \sim 10^{11.5} h^{-1}M_\odot$ at $z=0.5$), scatter in the mass estimates artificially inflates the number density by a factor of $\sim 3$ relative to the true mass function, and also distorts the shape near the characteristic ``knee'' mass. Despite these systematic effects, the recovered stellar mass functions from both the mock catalogue and real PAUS data show reasonable agreement with expectations at low to intermediate redshifts, with discrepancies increasing at higher redshift as the observed photometric bands shift to shorter rest-frame wavelengths.

The symbolic regression approach offers several important advantages over traditional methods. The explicit mathematical expressions reveal which observables drive the mass estimates and how, providing transparency that is lacking in both SED fitting and neural networks. Simple functional forms are more robust to observational noise than complex models. The expressions can be evaluated instantaneously and are easily portable to any analysis pipeline. Finally, the coefficients can be interpreted physically in terms of luminosity, colour, and $k$-correction effects. However, important limitations must be acknowledged. The expressions are optimized for the specific survey characteristics and galaxy formation model used for training, and transferability to other surveys requires careful validation. Massive blue galaxies and low-mass galaxies near the survey limit show larger systematic errors. Unlike SED fitting, the simple regression expressions do not provide native per-galaxy uncertainty estimates, though bootstrap or ensemble methods could address this. The expressions also inherit modelling assumptions from the training data, including dust attenuation prescriptions and stellar population synthesis models.

Several directions for future work are promising. Incorporating near-infrared photometry, which more directly traces stellar mass, could improve accuracy particularly for dusty and high-redshift galaxies. Training multiple models on different mock realizations could quantify systematic uncertainties and improve robustness. Developing separate expressions optimized for narrow redshift ranges might reduce systematic biases at high redshift. Methods for providing per-galaxy uncertainty estimates would make the approach more competitive with traditional SED fitting. The same methodology could be applied to estimate other physical properties such as star formation rates or dust masses. Finally, testing whether expressions trained on one survey can be calibrated for another with minimal additional data would be valuable for survey planning.

In conclusion, we have shown that symbolic regression provides a powerful, transparent, and computationally efficient approach to stellar mass estimation that complements traditional SED fitting methods. For applications requiring rapid analysis of large datasets or physical insight into mass estimation systematics, our approach offers significant advantages. The expressions we derive achieve accuracy comparable to SED fitting while revealing the physical relationships between observables and stellar mass in an explicit, interpretable form. For the bulk of the galaxy population in magnitude-limited surveys, simple linear expressions combining luminosity, colour, and redshift information can recover stellar masses with uncertainties similar to those from full SED modelling, but with vastly reduced computational cost and increased transparency.

\section*{Acknowledgements}
AK acknowledges the hospitality and support of Durham University during a short visit. 
CMB acknowledges support from the Science Technology Facilities Council (ST/X001075/1).
SK acknowledges a studentship from the Thai Government. 
GM acknowledges the support of the SKL of Displays and Opto-Electronics (project 2023 SKL ITC-PSKL12EG02 R1X27)
This work used the DiRAC@Durham facility managed by the Institute for Computational Cosmology on behalf of the STFC DiRAC HPC Facility (www.dirac.ac.uk). The equipment was funded by BEIS capital funding via STFC capital grants ST/K00042X/1, ST/P002293/1, ST/R002371/1 and ST/S002502/1, Durham University and STFC operations grant ST/R000832/1. DiRAC is part of the National e-Infrastructure.
JGB acknowledges support from the Spanish Research Project PID2021-123012NB-C43 [MICINN-FEDER], and the Centro de Excelencia Severo Ochoa Program CEX2020-001007-S at IFT.
 EG acknowledges support from the Spanish Ministerio de Ciencia e Innovacion (MICINN), project PID2021-128989NB. DNG acknowledge support from the European Research Council (ERC) under the European Union’s Horizon 2020 research and innovation program with Grant agreement No. 101053992. 
 This work was partially supported by the European Union's Horizon 2020 Research and Innovation program under the Maria Sklodowska-Curie grant agreement (No. 754510). 
 The PAU Survey is partially supported by MINECO under grants CSD2007-00060, AYA2015-71825, ESP2017-89838, PGC2018-094773, PGC2018-102021, PID2019-111317GB, SEV-2016-0588, SEV-2016-0597, MDM-2015-0509 and Juan de la Cierva fellowship and LACEGAL and EWC Marie Sklodowska-Curie grant No 101086388 and no.776247 with ERDF funds from the EU Horizon 2020 Programme, some of which include ERDF funds from the European Union. IEEC and IFAE are partially funded by the CERCA and Beatriu de Pinos program of the Generalitat de Catalunya. Funding for PAUS has also been provided by Durham University (via the ERC StG DEGAS-259586), ETH Zurich, Leiden University (via ERC StG ADULT-279396 and Netherlands Organisation for Scientific Research (NWO) Vici grant 639.043.512), University College London and from the European Union's Horizon 2020 research and innovation programme under the grant agreement No 776247 EWC. The PAU data center is hosted by the Port d'Informaci\'o Cient\'ifica (PIC), maintained through a collaboration of CIEMAT and IFAE, with additional support from Universitat Aut\`onoma de Barcelona and ERDF. The authors acknowledge the PIC services department team for their support and fruitful discussions.
\section*{Data Availability}
The primary outputs of this work are the symbolic regression equations. We used mock catalogues from \cite{Manzoni:2024}.



\bibliographystyle{mnras}
\bibliography{symbolic_main} 


\bsp	
\label{lastpage}
\end{document}